\documentclass[aps,prx,nobibnotes,twocolumn,notitlepage,superscriptaddress]{revtex4-1}
\usepackage{graphicx}
\usepackage{bm}
\usepackage{amsmath}
\usepackage{amssymb}
\usepackage{euscript}
\usepackage{verbatim}
\usepackage{fancyhdr}
\usepackage{wrapfig}
\usepackage{setspace}
\usepackage{xcolor}
\usepackage{amsfonts}
\usepackage{subfigure}
\usepackage{array}
\usepackage[colorlinks=true,linkcolor=blue,citecolor=blue]{hyperref}

\newcommand{\be}{\begin{equation}}
\newcommand{\ee}{\end{equation}}
\newcommand{\bea}{\begin{eqnarray}}
\newcommand{\eea}{\end{eqnarray}}

\newcommand{\rev}[1]{\textcolor{black}{#1}}

\begin{document}

%\begin{titlepage}

\title{Pretopological fractional excitations in the two-leg flux ladder}
\author{Marcello Calvanese Strinati}
\affiliation{Department of Physics, Bar-Ilan University, 52900 Ramat-Gan, Israel}

\author{Sharmistha Sahoo}
\affiliation{Department of Physics and Astronomy and Quantum Materials Institute, University of British Columbia, Vancouver, British Columbia, Canada V6T 1Z1}

\author{Kirill Shtengel}
\affiliation{Department of Physics and Astronomy, University of California, Riverside CA 92511, USA}

\author{Eran Sela}
\affiliation{Raymond and Beverly Sackler School of Physics and Astronomy, Tel-Aviv University, 69978 Tel-Aviv, Israel}
\affiliation{Department of Physics and Astronomy and Quantum Materials Institute, University of British Columbia, Vancouver, British Columbia, Canada V6T 1Z1}

\date{\today}
	
\begin{abstract}
%The emergence of non-Abelian anyons as effective particles in one-dimensional systems has been claimed to be limited by symmetry-based classification schemes to Majorana fermions. One route around is the search of such exotic particles on the edge of two-dimensional topological systems. Here we uncover another loophole in the rigorous arguments based on emergent symmetry of emergent non-topological degrees of freedom in non-homogenous systems. As the proof of concept, we show that protected parafermions can be obtained  on the edges of a one-dimensional chain of interacting bosons or fermions. Our construction is based on a pre-topological aspects of the thin torus limit.
Topological order, the hallmark of fractional quantum Hall states, is
primarily defined in terms of ground-state degeneracy on higher-genus
manifolds, e.g. the torus. We investigate analytically and numerically the
smooth crossover between this topological regime and the Tao-Thouless thin
torus quasi-1D limit. Using the wire-construction approach, we analyze an
emergent charge density wave (CDW) signifying the break-down of topological
order, and relate its phase shifts to Wilson loop operators.
% governing topological order in the 2D limit.
The CDW amplitude decreases exponentially with the torus circumference once
it exceeds the transverse correlation length controllable by the inter-wire
coupling. By means of numerical simulations based on the matrix product states
(MPS) formalism, we explore the extreme quasi-1D limit in a two-leg flux
ladder and present a simple recipe for probing fractional charge excitations
in the $\nu=1/2$ Laughlin-like state of hard-core bosons. We discuss the
possibility of realizing this construction
%measurement of CDWs and fractional charges
in cold-atom experiments. We also address the implications of our findings to
the possibility of producing non-Abelian zero modes. As known from rigorous
no-go theorems, topological protection for exotic zero modes such as
parafermions cannot exist in 1D fermionic systems and the associated
degeneracy cannot be robust. Our theory of the 1D-2D crossover allows to
calculate the splitting of the degeneracy, which vanishes exponentially with
the number of wires, similarly to the CDW amplitude.
\end{abstract}
	
\maketitle

\section{Introduction and main results}
Topologically ordered phases of matter have attracted significant attention
because of their potential utility for quantum computation. Indeed, because of
the intrinsically nonlocal nature of their order, topological phases can host
anyonic excitations that are robust to any local perturbation. Because of that,
such phases can serve as a platform for fault-tolerant quantum computation,
also known as topological quantum
computation~\cite{kitaev2003fault,doi:10.1142/3724,quant-ph/0004010,AVERIN200125,quant-ph/0101025,sarma2006quantumcomputation,nayak2008non}.
Finding physical systems supporting topological order that can be used for
quantum computation purposes has been a challenging task throughout the last
decades. Fractional quantum Hall (FQH) states could provide one such platform~\cite{quant-ph/0101025,sarma2006quantumcomputation,nayak2008non,ezawa2008quantum}.

First observed experimentally in 1982 in strongly-interacting two-dimensional
(2D) electron gases~\cite{PhysRevLett.48.1559}, FQH states are characterized by
the existence of exotic fractionally charged
excitations~\cite{laughlin1983anomalous} with anyonic
statistics~\cite{Halperin1984,Arovas1984}. A striking consequence of both
fractional statistics and fractional charge of such excitations is the so-called
topological ground-state degeneracy on the
torus~\cite{wen1990ground,Oshikawa2006}. Such degeneracy depends only on the
type of topological order and the genus of a surface; it can not be probed by
any local measurements. Consequently, it is used to \emph{define} the very
notion of topological order~\cite{wen1990ground}.

On the contrary, the state proposed for the fractional quantum Hall effect by
Tao and Thouless in 1983~\cite{tao1983fractional}, shortly after Laughlin's
work~\cite{laughlin1983anomalous}, displays charge density wave (CDW) order,
which breaks translation symmetry. The number of the CDW ground states on the
torus matches the aforementioned topological degeneracy of the Laughlin state
(which is in turn related to the filling factor $\nu$), and fractional charge
excitations are given by domain walls between the degenerate ground
states~\cite{PhysRevB.28.2264,PhysRevB.30.1069,PhysRevB.32.2617}. It has since
been established that the Tao and Thouless state is the ground state in the
limit when the small circumference of the torus (which we call $L_y$) is comparable
to the magnetic length~\cite{Bergholtz2006a,Bergholtz2006b,Seidel2006,Seidel2007a,bergholtz2008quantum}, and that it is adiabatically connected to the Laughlin
state (see Refs.~\cite{hansson2009tao,kjall2018matrix} and references therein).

The goal of this paper is to provide an analytical and numerical description of this crossover between the thin torus quasi one-dimensional (1D) limit, and the 2D Laughlin limit of the FQH effect \rev{at zero temperature} in systems of many coupled wires subjected to effective magnetic fluxes (referred to as \emph{flux ladders}). Such crossover can be obtained by resorting to the wire-construction approach of the Laughlin state discussed in Refs.~\cite{PhysRevLett.88.036401,teo2014luttinger}. The reason for relying on the coupled-wire approach stems from the fact that the realization of flux ladders is currently at the experimental and numerical reach, thanks to the amazing progresses in the field of ultra-cold atoms, which provide the toolbox for creating and probing synthetic matter using atomic gases in optical lattices~\cite{bloch2005ultracoldatoms,doi:10.1080/00018730701223200,RevModPhys.80.885,RevModPhys.83.1523,goldman2016topological}, and the realization of \emph{ad-hoc} numerical algorithms based on the density-matrix-renormalization-group (DMRG)~\cite{PhysRevLett.69.2863,RevModPhys.77.259} or matrix-product-state (MPS)~\cite{SCHOLLWOCK201196} formalism.

Several interesting properties have been highlighted by a number of works, for
both
bosonic~\cite{PhysRevB.64.144515,PhysRevA.85.041602,petrescu2013bosonic,PhysRevB.87.174501,PhysRevA.89.063617,1367-2630-16-7-073005,DiDio2015,PhysRevB.91.140406,PhysRevB.92.060506,1367-2630-17-9-092001,PhysRevA.92.053623,PhysRevLett.115.190402,PhysRevA.94.063628,1367-2630-18-5-055017,Strinati_2018,PhysRevA.98.033605,PhysRevA.99.053601}
and fermionic~\cite{PhysRevB.71.161101,1367-2630-17-10-105001,ncomms9134,1367-2630-18-3-035010,PhysRevA.95.063612,PhysRevLett.118.230402,PhysRevA.93.013604,PhysRevA.93.023608,Haller_2018}
flux ladders. Importantly, it was shown that flux ladders, in the quasi-1D limit, can
host states that share fundamental properties with 2D FQH
states~\cite{petrescu2015chiral,cornfeld2015chiral,strinati2017laughlin,petrescu2017precursor},
and that can be directly tested in current cold-atom experiments as well as
DMRG or MPS simulations~\cite{strinati2017laughlin,petrescu2017precursor}.
Specifically, quantum Hall states with finite transverse dimension $N_w$, can
be realized using 1D cold atom lattices, by combining synthetic
dimensions~\cite{PhysRevLett.108.133001,goldman2015fourdimensional,goldman2017synthetic}
with synthetic gauge fields~\cite{RevModPhys.83.1523}. Indeed, as was recently
demonstrated in two independent
experiments~\cite{mancini2015observation,stuhl2015visualizing}, one can produce
a quantum Hall ribbon with edge states. In this case, Raman lasers were used to
drive transitions between three atomic states, simulating a three-leg ladder.
Furthermore, synthetic quantum Hall stripes can be effectively ``rolled'' into
thin cylinders~\cite{celi2014synthetic,PhysRevLett.122.065303}.

A variety of other physical systems can realize 2D topological states using 1D systems with synthetic dimensions and gauge fields, such as integrated photonic systems~
%a 1D array of optical cavities with a synthetic dimension
\cite{luo2015quantum} with orbital angular momentum of light playing the role
of synthetic dimension, or even frequency
modes~\cite{ozawa2016synthetic,yuan2016photonic}, for a review see
Ref.~\cite{RevModPhys.91.015006}. A key issue is the prospect of strong
particle-particle interactions, which in optics are mediated by strong
nonlinearities, to realize topologically nontrivial strongly correlated states.
%Common to cold atoms and optical systems, the fact that interactions are often nonlocal in synthetic dimension~\cite{celi2014synthetic} is beneficial for few-leg ladders where it is not too non nonlocal.

From a more theoretical point of view, topological degeneracy of FQH states
(even Abelian ones, such as those considered in this paper) can be used to
generate \emph{non-Abelian} topological defects, genons, which effectively
change the genus of the underlying surface~\footnote{In non-Abelian states,
additional topological degeneracy is associated with excitations (and is
independent of the genus), which results in the extensive degeneracy of excited
states that can in turn be used to store and manipulate quantum
information~\cite{nayak2008non}.}. %These fractionalized objects can emerge on interfaces between different topological phases, or vortices in certain 2D topological systems. A degeneracy is then associated with the \emph{fusion outcome} of pairs of non-Abelian anyons.
This possibility is of particular interest to us. While a rich variety of
non-Abelian anyons may potentially exist in FQH and other 2D topological
phases~\cite{nayak2008non}, few of those states have been accessed
experimentally to date, and none can thus be utilized for quantum
computation. Meantime, much of the recent progress has been in using quasi-1D
systems to produce one type of non-Abelian objects, Majorana zero
modes~\cite{Lutchyn2018}~\footnote{We will avoid referring to non-Abelian
modes in quasi-1D systems as anyons for two reasons. Firstly, they are not
proper quasiparticles, i.e. they are not low-energy excitations in a
topological state. Instead, they are zero modes bound to some sorts of
topological defects which themselves are very high-energy. Secondly, their
defining feature, the braiding statistics, is not well defined in 1D. While it
is possible to devise schemes for braiding such
objects~\cite{Alicea2011a,Clarke2011a,Sau2011a}, such schemes invariably take
one outside of a quasi-1D setting and are not a focus of this study.}. All
attempts to come up with more exotic types of non-Abelian zero modes in 1D
interacting  fermionic systems have run into seemingly restrictive no-go
theorems~\cite{fidkowski2011topological,turner2011topological,bultinck2017fermionic}.
% result in no-go theorems for anyons beyond Majorana fermions on 1D. Such exotic anyons can be realized in 1D, however in this case they are not topologically protected, i.e. one can physically distinguish their various fusion outcomes via local observables even in the limit of infintie separation between anyons in contrast to the 2D case~\cite{chen2016tunable}.
%While several works studied intrinsic properties of parafermion
%chains~\cite{fendley2012parafermionic,jermyn2014stability,fendley2014free,mong2014parafermionic,stoudenmire2015assembling,alicea2016topological,iemini2017topological,mazza2018nontopological,meichanetzidis2018free,mazza2018energy,rossini2019anyonic},
%the no-go theorems raise the impossibility to realize such effective particles
%from microscopic fermionic or bosonic Hamiltonians in 1D.
One way to circumvent such restrictions is to use 1D edge states of 2D
topologically ordered
systems~\cite{barkeshli2012topological,barkeshli2013twist,barkeshli2014synthetic,Clarke2013a,Lindner2012,Cheng2012,vaezi2013fractional,klinovaja2014j,barkeshli2015generalized,orth2015non,vinkler2017z,zhang2014time}.
However, in strictly 1D systems, the degeneracy associated with non-Abelian
zero modes may always be removed by local perturbations. For example, the intrinsic properties of parafermion zero modes in 1D have been studied in several works~\cite{fendley2012parafermionic,jermyn2014stability,fendley2014free,mong2014parafermionic,PhysRevB.90.195101,stoudenmire2015assembling,alicea2016topological,iemini2017topological,mazza2018nontopological,meichanetzidis2018free,mazza2018energy,rossini2019anyonic}, and furthermore, parafermion-like zero modes have been obtained in a number of 1D
proposals~\cite{oreg2014fractional,klinovaja2014parafermions,calzona2018z4}. However, both the zero-energy nature of such modes and the the associated
ground state degeneracy (in the presence of such modes) are unstable against
local perturbations, as has been explicitly checked
~\cite{oreg2014fractional,calzona2018z4}.

Here, we establish the connection between the loss of topological protection for
such non-Abelian modes in quasi-1D FQH systems with the emergence of a CDW in
the thin torus limit. Such modes gradually become topologically protected upon
increasing the number of 1D wires, $N_w$. Any local observable distinguishing
their ground states is suppressed exponentially in $N_w$. Hence, these
non-Abelian zero modes can be effectively realized in 1D systems with a finite
width. Our key finding is that while the no-go theorems predict the absence of
topological protection in 1D for any zero modes that are more exotic than
Majorana zero modes, the energy splitting between their ground states can be made
vanishingly small, along with the CDW amplitude.

\begin{figure} [t]
	\centering	
	\includegraphics[width=8.5cm]{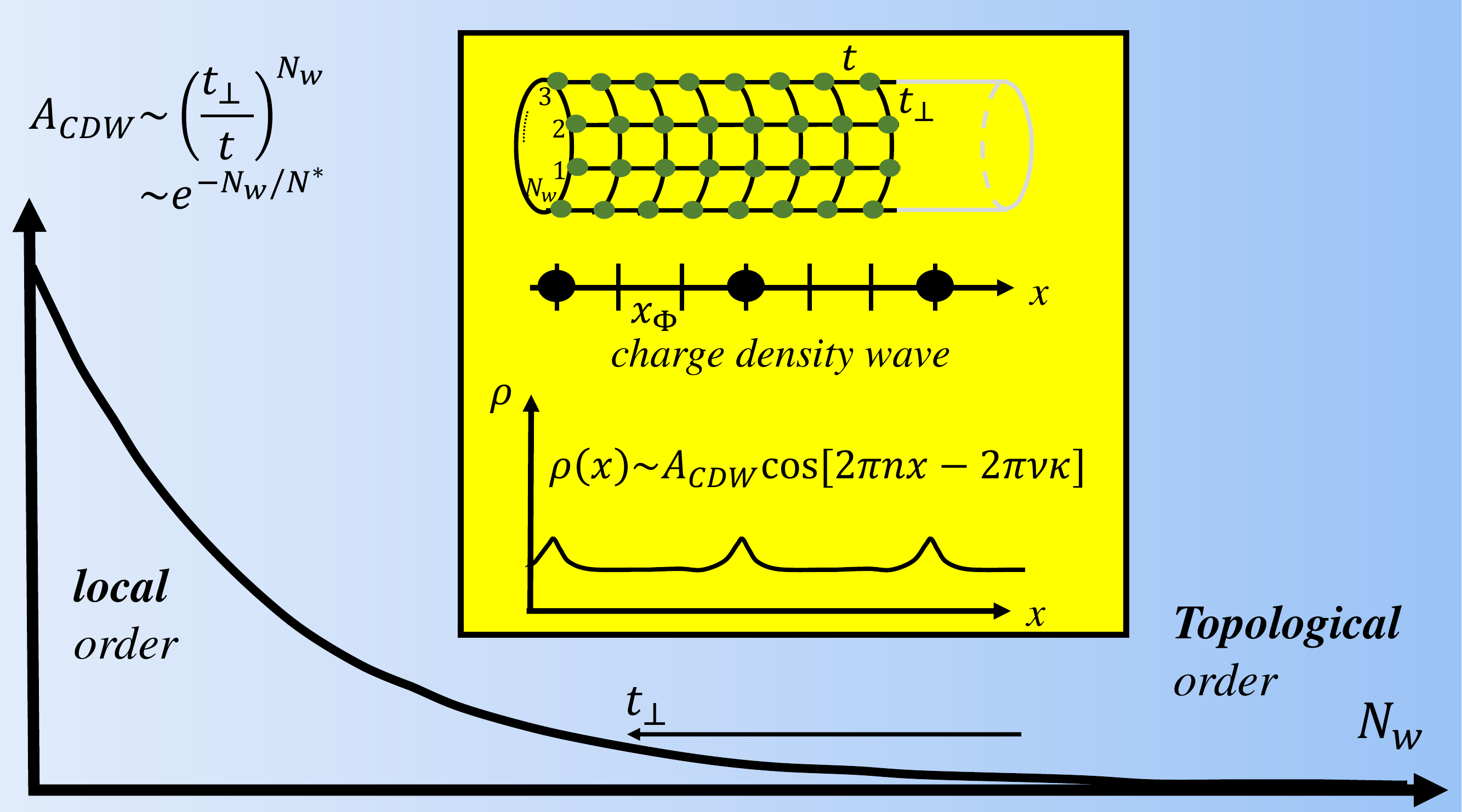}
	\caption{Phase diagram of a FQH state on an $N_w$-leg ladder. The translation symmetry by one magnetic translation vector $x_\Phi = 2 \pi /(\Phi N_w)$ is broken by a charge density wave $\rho(x) = A_{\rm CDW}^{(N_w)}\cos(2 \pi n x - 2 \pi \nu \kappa)$, whose amplitude $A_{\rm CDW}^{(N_w)}$ vanishes in the thick cylinder limit, or in the anisotropic limit $t_\perp/t \to 0$. Here, $\nu$ is the filling factor, $n=N_w \rho_0$ is the density of the quasi-1D system, where $\rho_0$ is the density per chain, and $\kappa \in \mathbb{N}$.}
	\label{fig:1}
\end{figure}

The rest of this paper is organized as follows. In Sec.~\ref{se:1D2D}, we
construct the Laughlin state at filling factor $\nu$ on a system of $N_w$ wires
weakly coupled by tunneling $t_\perp$ and rolled into a cylinder or torus, see
Fig.~\ref{fig:1}. In the Tao-Thouless limit of small $N_w$, a CDW forms with
amplitude $A_{\rm CDW}^{(N_w)}$. Using the wire-construction approach, we show
that it originates from $N_w$-th order perturbation theory in $t_\perp$,
implying that the local order parameter $A_{\rm CDW}^{(N_w)}$ is actually nonlocal in
the transverse direction. As follows from general
arguments~\cite{kitaev2003fault}, in the topological limit of large $N_w$, the
CDW amplitude should be exponentially small. In the few-leg ladders on which we
focus, this substantiates that the CDW degeneracies are the pre-topological
limit of the Laughlin state.

In Sec.~\ref{se:2legladder}, we discuss how one can realize an effective thin cylinder in the extreme 1D limit with width $N_w=2$, i.e., the two-leg flux ladder. In this limit, excitations are domain walls, and we identify their fractional charge via numerical simulations based on MPS. We focus on the FQH Laughlin-like state of bosons at filling factor $\nu=1/2$, by using the same numerical scheme as in Ref.~\cite{strinati2017laughlin}. The advantage of focusing on the $\nu=1/2$ Laughlin-like state is that stabilizing this fractional state requires only on-site interactions, as opposed to smaller $\nu$, e.g. $\nu=1/3$ for fermions or $\nu=1/4$ for bosons, which require longer range interactions~\cite{cornfeld2015chiral}, which makes their numerical simulation more demanding. In the $\nu=1/2$ case, %, since we can only simulate a ladder with finite longitudinal dimension (i.e., with a finite number of sites $L$ per leg) and open boundary conditions (OBC),
we find two quasi-degeneretate CDW states. We simulate domain-wall excitations and show that they have charge $1/2$. By resorting to our numerical analysis, and from the wire construction, we conclude that such fractional domain-wall excitations are the pre-topological limit of the Laughlin quasiparticles.

The purpose of Sec.~\ref{se:anyons1D} is to point out that, similar to the
emergence of a CDW in the thin torus limit of the FQH effect, 1D systems
hosting exotic zero modes such as parafermions undergo a 1D-2D  crossover (from
non-topological to topological) that can be controlled by the number of 1D
wires $N_w$.  We finally present our conclusions in Sec.~\ref{sec:conclusions}.

\begin{figure}[t]
\centering
\includegraphics[width=8.4cm]{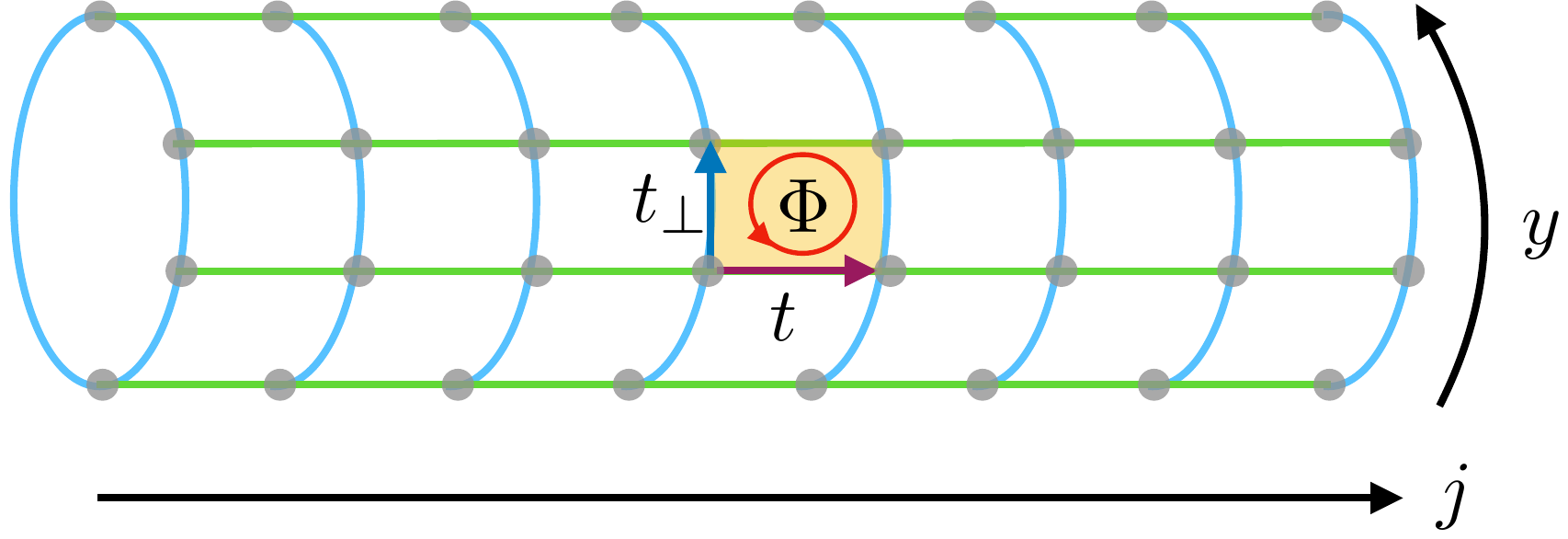}
\caption{Scheme of the $N_w$-leg ladder closed on a cylinder. The system consists of $N_w$ chains (or legs or wires) of $L$ sites each. Particles on the lattice can hop between nearest-neighbour lattice sites (grey dots) along each chain, with tunneling amplitide $t$ (purple arrow), or between nearest-neighbour chains with tunneling amplitude $t_\perp$ (blue arrow). When encircling a closed loop (red arrow) delimited by four nearest-neighbour lattice sites (a \emph{plaquette}, yellow area) a phase factor equal to the gauge flux per plaquette $\Phi$ is gained. The longitudinal and transverse \rev{dimensions} are denoted by $j$ and $y$, respectively, \rev{where $j=1,\ldots,L$ and $y=1,\ldots,N_w$, and the set of $N_w$ sites at the same $j$ form a rung of the ladder}.}
\label{fig:cylinderscheme}
\end{figure}

\section{Wire construction: charge density wave and 1D-2D crossover}
\label{se:1D2D}
We open our discussion on the 1D-2D crossover by focusing on the CDW amplitude of the Laughlin state realized on the $N_w$-leg ladder. \rev{We choose to consider a geometry of a cylinder rather than a torus, since it is more realistic in experimental and numerical contexts~\cite{PhysRevA.93.013604,PhysRevA.99.033603}. In 2D, the fate of putting a FQH state on a torus or on a cylinder is different, since the latter has edges. In the present section, however, even if not explicitly stated, we consider an infinite cylinder, and ignore its edges. We remark that these edges of a finite cylinder will have a role in the numerical simulations in Sec.~\ref{se:2legladder} relevant for experiments, on which we will comment later.}

\subsection{Model}
We consider bosonic or fermionic particles hopping on a cylinder, i.e. on an $N_w$-leg ladder with periodic boundary conditions (PBC) \rev{along the transverse dimension $y$, and open boundary conditions (OBC) along the longitudinal dimension $j$}, see Figs.~\ref{fig:1} and~\ref{fig:cylinderscheme}. Such system is modelled by the following Hamiltonian:
\begin{eqnarray}
\hat H&=&-t\sum_{j=1}^{L-1}\,\sum_{y=1}^{N_w}\left(\hat b^\dag_{j,y}\hat b_{j+1,y}+{\rm H.c.} \right)\nonumber\\
&&+t_\perp\sum_{j=1}^{L}\sum_{y=1}^{N_w}\left(e^{i\Phi j}\,\hat b^\dag_{j,y+1}\hat b_{j,y}+{\rm H.c.}\right)+\hat H_{\rm int} \,\, .
\label{eq:hamiltonianfermionctwolegladder}
\end{eqnarray}
Here, $\hat b_{j,y}$ ($\hat b^\dag_{j,y}$) represents the annihilation (creation) operator of a boson or fermion on site $j$ and on the leg $y=1,2, \dots, N_w$; $t$ and $t_\perp$ are the intra-leg and inter-leg hopping parameters, respectively, $\Phi$ is the gauge flux per plaquette (see Fig.~\ref{fig:cylinderscheme}). \rev{In the Hamiltonian in Eq.~\eqref{eq:hamiltonianfermionctwolegladder}, $L$ denotes the number of lattice sites in the longitudinal dimension $j$, which we take $L\rightarrow\infty$ since we consider an infinite-cylinder. Furthermore,} for PBC \rev{along the transverse dimension $y$}, $\hat b_{j,N_w+1}=\hat b_{j,1}$. The term $\hat H_{\rm int}$ accounts for density-density interactions that we will specify in Sec.~\ref{se:2legladder}, which is needed in order to stabilize FQH states. One defines the quantum Hall filling factor $\nu = 2\pi\rho_0/\Phi$, where $\rho_0 = \langle \hat{b}^\dagger_{j,y} \hat{b}_{j,y} \rangle$ is the average density per site, on the wire~$y$. Accordingly, we define the total density of the quasi-1D system as $n = N_w \rho_0$. In the following, we specialize to Laughlin states with $\nu =1/(2q)$, where $q>0$ is an integer for bosons or an half-integer for fermions.

Manifestations of the FQH  effect on this $N_w$-leg ladder has been an active topic of research, both for the OBC case with edge states~\cite{petrescu2013bosonic,petrescu2015chiral,cornfeld2015chiral,strinati2017laughlin,petrescu2017precursor}, and for PBC corresponding to a cylinder. Specifically, in the latter case, it was shown that the Laughlin pumping takes place effectively for thin cylinders~\cite{PhysRevLett.118.230402} which physically operates as a drifting CDW pattern.

In this section, we focus on the latter, non-topological feature of the FQH state on a cylinder, namely, the appearance of a CDW, which can be in one out of $2q$ states (see inset of Fig.~\ref{fig:1}) characterized by the same CDW pattern but relatively shifted in real space by an amount that is a multiple of $x_\Phi = \nu /n=2 \pi/ (\Phi N_w)$. Explicitly, we will show that the leading CDW harmonic has the form
\begin{equation}
\label{eq:CDWbasic}
\rho(x) \sim \rho_0+ A_{\rm CDW}^{(N_w)} \cos(2 \pi n x - 2 \pi \nu \kappa).
\end{equation}
The first argument of the cosine suggests a Wigner crystal whose period is dictated by the particle density. The second is a discrete shift of the CDW, with $\kappa \in \mathbb{N}$, which takes only one out of $\nu^{-1}$ ground states.
%The latter reminds us of the fact that in the presence of a magnetic field the system only has a discrete translation symmetry by $x_\Phi =2 \pi / (\Phi N_w)$.
Such $2q$-fold degeneracy is consistent with the number of FQH ground states on a torus. We now compute the amplitude of this charge density wave $A_{\rm CDW}^{(N_w)}$ and demonstrate that it originates from $N_w$-th order perturbation theory in the inter-chain hopping amplitude $t_\perp$.

\subsection{Low-energy approach}
We use a continuum theory to compute the density using bosonization based on the wire-construction approach. The lattice operator $\hat b_{j,y}$ is replaced by a field operator $\Psi_y(x)$, which is expanded in terms of a charge (or density) field $\phi_y$ and a phase field $\theta_y$ as~\cite{giamarchi2003quantum}
\begin{equation}
\begin{array}{ll}
\displaystyle{\hat{b}^\dagger_{j,y} \sim \Psi_y^\dagger (x) = \sum_p \psi^\dagger_{y,p}(x)}\\\\
 \psi^\dagger_{y,p}(x) = \alpha_{p,y}\,e^{i p [2 \pi \rho_0 x - 2 \phi_y(x)]} e^{- i \theta_y (x)} \,\, ,
\end{array}
\label{eq:expansionforthefieldsbosonization}
\end{equation}
where $p$ is an integer for bosons and half integer for fermions. The charge and phase fields obey canonical commutation relations $\left[\partial_x\phi_{y}(x), \theta_{y'}(x')\right] = -i\pi \,\delta_{y,y'}\, \delta(x~-x')$. In Eq.~\eqref{eq:expansionforthefieldsbosonization}, $\{\alpha_{p,y}\}$ are non-universal expansion coefficients that depend on the microscopic details of the model~\cite{RevModPhys.83.1405} (they do not depend on the wire index $y$, this index is kept for clarity). We have set the lattice constant to unity, $a=1$. Likewise, the density field at wire $y$ has the expansion~\cite{RevModPhys.83.1405}
\begin{equation}
\label{eq:rhoexp}
\rho_y (x) = \Psi_y^\dagger(x)\Psi_y(x) = \rho_0 - \frac{1}{\pi}\, \partial_x \phi_y(x)+\sum_{p\,\in\,\mathbb{Z}/\{0\} } \rho_y^{(p)}(x) \,\, ,
\end{equation}
where $ \rho_y^{(p)}(x)=\beta_{p,y}\,e^{i p [2 \pi \rho_0 x - 2 \phi_y(x)]}$ for some non-universal expansion coefficients $\{\beta_{p,y}\}$. We stress that, in the expansion of the density in Eq.~\eqref{eq:rhoexp}, $p$ is an integer for both bosons and fermions. Within this framework, we write the Hamiltonian of the system as
\begin{equation}
\hat{H} = \hat{H}_0 + \hat{H}_{\rm int} + \hat{H}_\perp \,\, ,
\label{eq:totalhamiltonianbosonization}
\end{equation}
where $\hat{H}_0$ and $\hat{H}_{\rm int}$ are the continuum versions of the intra-leg hopping and interaction terms in Eq.~(\ref{eq:hamiltonianfermionctwolegladder}), respectively, describing a gapless Luttinger liquid, whereas $\hat H_\perp$ describes the continuum version of the inter-leg hopping term accounting for the gauge flux in Eq.~\eqref{eq:hamiltonianfermionctwolegladder}, whose bosonized form reads
\begin{eqnarray}
\label{eq:Oppp}
 \hat{H}_\perp &=& t_\perp  \int dx \sum_{y=1}^{N_w} \Psi_{y+1}^\dagger(x)\Psi_{y}(x)  e^{i \Phi x} +{\rm H.c.} \nonumber \\
 &=&t_\perp \int dx\,e^{i \Phi x}   \sum_{y=1}^{N_w} \, \sum_{p  , p'}\,\alpha^*_{p,y}\,\alpha_{p',y+1}\nonumber\\
 &&\hspace{1cm}\times\,e^{ -i (p-p') 2 \pi \rho_0 x} \,\mathcal{O}_{p p'}^{y \to y+1}+{\rm H.c.} \,\, .
\end{eqnarray}
Here
\begin{equation}
\mathcal{O}_{p p'}^{y \to y+1}  \sim  e^{i[\theta_{y} - \theta_{y+1} + 2 (p\,\phi_{y} -p' \,\phi_{y+1} )]} \,\,
\label{eq:linktunnelingoperator}
\end{equation}
is the link tunneling operator between the legs $y$ and $y+1$. In the following, in order to ease the notation, we introduce the non-universal coefficient $C_{p,p'}^{y,y+1}=\alpha^*_{p,y}\,\alpha_{p',y+1}$. Therefore, Eq.~\eqref{eq:Oppp} can be written as
\begin{eqnarray}
\label{eq:Hperp}
\hat{H}_{\perp} &=&t_\perp \int dx \sum_{y=1}^{N_w}\,\sum_{p,p'}C_{p,p'}^{y,y+1}\,\cos[ \theta_{y} - \theta_{y+1} \nonumber\\
&&\hspace{0.1cm}+2 (p\,\phi_y -p' \,\phi_{y+1} )- (p-p') 2 \pi \rho_0 x+ \Phi x] \,\, .
\end{eqnarray}
In order to describe fluctuations within the FQH phase, we separate the various terms in Eq.~\eqref{eq:Hperp} as $\hat{H}_{\perp} =~ \hat{H}_{\rm FQH} + \delta \hat{H}$, where $\hat{H}_{\rm FQH}$ corresponds to the non-oscillating terms $(p - p')2 \pi \rho_0  = \Phi $ with $p'=-p$ in the sum over $p$ and $p'$, and for a fixed value of $p\equiv q>0$ that determines the fractional filling factor, i.e., $\nu~=~2\pi\rho_0/\Phi =~(2q)^{-1}$~\cite{cornfeld2015chiral}, whereas $\delta\hat H$ contains all the other combinations of $p$ and $p'$:
\begin{eqnarray}
\delta \hat{H}&=& t_\perp \int dx \, e^{i \Phi x} \ \sum_{y=1}^{N_w}  \,\sum_{p  , p'}' \,C_{p,p'}^{y,y+1}\,e^{ -i (p-p') 2 \pi \rho_0 x} \nonumber\\
&&\hspace{3cm}\times\, \mathcal{O}_{p p'}^{y \to y+1}+{\rm H.c.} \,\, ,
\end{eqnarray}
where the primed sum $ \sum'_{p  , p'}$ does not contain the FQH operator $p= - p ' = q=(2 \nu)^{-1}$. Therefore, the total Hamiltonian in Eq.~\eqref{eq:totalhamiltonianbosonization} is recast as $\hat H=\hat H_{0}+\hat H_{\rm int}+ \hat H_{\rm FQH}+\delta\hat H$, where we treat $\delta\hat H$ as a perturbation.
In detail, $\hat H_{0}+\hat H_{\rm int}+ \hat H_{\rm FQH}$ consists of $N_w$ decoupled sine-Gordon models~\cite{teo2014luttinger}:
$\hat H_{0}+\hat H_{\rm int}$ map to $N_w$ Luttinger liquids, characterized below just by a velocity $v$. The Luttinger liquids are gapped out by the cosine potentials $\hat H_{\rm FQH}$. Indeed, it is convenient to introduce the gapped link fields $\tilde{\phi}_{y+\frac{1}{2}}$ and their strongly fluctuating conjugate fields $\tilde{\theta}_{y+\frac{1}{2}}$ as
\begin{equation}
\begin{array}{l}
2 \,\tilde{\phi}_{y+\frac{1}{2}} =  \theta_y - \theta_{y+1} + 2q\, (\phi_y + \phi_{y+1})  \\\\
2 \,\tilde{\theta}_{y+\frac{1}{2}} =  \theta_y + \theta_{y+1} + 2q\, (\phi_y - \phi_{y+1}) \,\, ,
\end{array}
\label{eq:fieldstildephiandconjugatedfield}
\end{equation}
so that $\hat{H}_{\rm FQH} \sim - \sum_{y=1}^{N_w} \cos (2 \tilde{\phi}_{y + \frac{1}{2}})$.  The fields in Eq.~\eqref{eq:fieldstildephiandconjugatedfield} satisfy the commutation relations $[\partial_x \tilde{\phi}_{y+\frac{1}{2}}(x), \tilde{\theta}_{y'+\frac{1}{2}}(x')]=2 q \left[\partial_x \phi_{y}(x), \theta_{y'}(x')\right]$. These cosine perturbations are assumed to be relevant and flow to strong coupling.  We denote the gap created by these cosine potentials $\Delta_{{\rm{gap}}}$. It leads to a correlation length $\xi = v/\Delta_{{\rm{gap}}}$.   On the other hand, the cosine operators in $\delta\hat H$ and oscillating terms will be treated perturbatively.

\subsection{Computation of the density}
Having introduced the required notation, we now compute the density $\langle \rho_y(x) \rangle$  by proceeding with a perturbative approach in terms of $\delta\hat H$. The density $\rho_y(x)$, which is expressed in terms of the original fields $\phi_y(x)$ and $\theta_y(x)$, is re-expressed in terms of the link fields and their conjugated fields by the inverse transformation of Eq.~\eqref{eq:fieldstildephiandconjugatedfield}:
\begin{equation}
\begin{array}{l}
\label{transfREV}
4q\, \phi_y =\tilde{\phi}_{y-\frac{1}{2} } - \tilde{\theta}_{y-\frac{1}{2} } + \tilde{\phi}_{y+\frac{1}{2} } + \tilde{\theta}_{y+\frac{1}{2} } \\\\
2\, \theta_y =  - \tilde{\phi}_{y-\frac{1}{2} } + \tilde{\theta}_{y-\frac{1}{2} } + \tilde{\phi}_{y+\frac{1}{2} } + \tilde{\theta}_{y+\frac{1}{2} } \,\, .
\end{array}
\end{equation}
In addition to pinned fields $\tilde{\phi}_{y\pm \frac{1}{2}}$, we see that the density field  $\phi_y$ contains the combination of $\tilde{\theta}_{y-\frac{1}{2}} - \tilde{\theta}_{y+\frac{1}{2}}$ of fluctuating fields. Therefore, prior to considering the effect of the perturbation $\delta\hat H$, we notice that $\langle e^{-2 i \phi_y} \rangle_0=0$, where the subscript ``$0$'' denotes that the expectation value is computed on the ground state of $\hat H$ with $\delta\hat H=0$. Hence, in this limit, the oscillating part of the density that contains information on the CDW vanishes, $\langle \rho_y \rangle_0= {\rm const}$.

On the other hand, consider a product of the density operators over all the wires $\prod_y e^{-2 i \phi_y} $. The telescopic series of fluctuating fields, using Eq.~(\ref{transfREV}),  yields a finite expectation value. Keeping the leading $p=1$ harmonic in Eq.~\eqref{eq:rhoexp} %, and neglecting constants arising from $\rho_0$,
we have
\begin{eqnarray}
\label{eq:productrho}
\left\langle \prod_{y=1}^{N_w} \rho_y^{(1)}(x)\right \rangle &=& \left(\prod_{y=1}^{N_w}\beta_{1,y}\right) e^{2 \pi i n x} \left\langle \prod_{y=1}^{N_w} e^{-2 i \phi_y}\right \rangle + {\rm H.c.}\nonumber\\
&=& 2 \left(\prod_{y=1}^{N_w}\beta_{1,y}\right)  \cos(2 \pi n x - 2 \pi \nu \kappa) \,\, ,
\end{eqnarray}
where, from Eq.~\eqref{transfREV}, the integer $\kappa$ is determined from
\be
\label{eq:kappa}
e^{-2 i\pi \nu \kappa} = \left\langle e^{-2 i \sum_y \phi_y} \right\rangle = \left\langle e^{-2 i \nu \sum_y \tilde{\phi}_{y\rev{+\frac{1}{2}}}} \right\rangle.
\ee
We will shortly show that, when computing the density at a specific wire perturbatively in $ \delta \hat{H} \propto t_{\perp}$, precisely this loop operator $ e^{-2 i \sum_y \phi_y} $ is generated in $N_w$-th order perturbation theory and yields the desired CDW in Eq.~(\ref{eq:CDWbasic}). Since it is a function of the pinned fields, it takes discrete values, reflecting a finite number of ground states. Specifically, $\kappa$ is an integer defined modulo $\nu^{-1}=~2q$. %This expression for $\kappa$ exemplifies the fact that while in the wire contruction approach each individual link field $\tilde{\phi}_{y+\frac{1}{2}}$ is pinned to an independent integer-value, only the sum of these integers modulo $2q$ is physical, yielding $2q$ distinguishable states.

In fact, the loop operator inside the expectation value in Eq.~(\ref{eq:kappa}) can be identified with an operator that transports a quasiparticle around the cylinder~\cite{teo2014luttinger}. These loop operators, known as Wilson loops, are crucial to understand the degeneracy of the FQH state on the torus~\cite{wen1990ground}. To be explicit, one can write a general  Wilson loop operator associated with a rectangular loop using the wire construction approach~\cite{teo2014luttinger,sagi2015imprint,gorohovsky2015chiral,santos2015fractional} as
\bea
\label{WLdetail}
W(\Box)=W^{x_1}_{y_2 \to y_1}W^{y_2}_{x_2 \to x_1}W^{x_2}_{y_1 \to y_2}W^{y_1}_{x_1 \to x_2} \,\, ,
\eea
where each factor transports a quasiparticle along a finite segment, with vertical segments $W_{y_1 \to y_2}^x = \prod_{y=y_1+1}^{y_2}e^{-2i \phi_y(x)}=\prod_{y=y_1+1}^{y_2}e^{-\frac{i}{2q} (\tilde{\phi}_{y-\frac{1}{2} } - \tilde{\theta}_{y-\frac{1}{2} } + \tilde{\phi}_{y+\frac{1}{2} } + \tilde{\theta}_{y+\frac{1}{2} }) }$ where the pinned fields give a constant phase factor, and horizontal segments
$W_{x_1 \to x_2}^y = e^{-\frac{i}{2q}\int_{x_1}^{x_2} \partial_x \tilde{\theta}_{y+\frac{1}{2}}}$~\cite{teo2014luttinger}. The nontrivial algebra satisfied by Wilson loops for non-contractible loops, e.g. $W(a) W(b) = W(b) W(a) \,e^{i\,2 \pi /(2q)}$ for the two nontrivial loops $a$ and $b$ on a torus, implies a $2q$ degeneracy~\cite{wen1990ground}.

%But for a thin cylinder, we can see from Eq.~(\ref{eq:kappa}) that the phase $\kappa$ of the CDW measures precisely the eigenvalue of the Wilson loop $c$ winding around the cylinder.
%Thus, we conclude that the phase of the CDW is associated with a loop operator, i.e., a product of density operators around the cylinder.
%

We now proceed with the perturbative expansion of the leading oscillating part of the density [i.e., $p=1$ in Eq.~\eqref{eq:rhoexp}] at a specific leg, which we denote by $y_0$:
\begin{eqnarray}
\left\langle \rho^{(1)}_{y_0}(x)\right\rangle \!& = &\! \frac{1}{Z} \int \mathcal{D}\phi\,\mathcal{D}\theta \, \rho^{(1)}_{y_0}(x) \nonumber\\
&&\!\times\, e^{ \int d\tau \int dx \,\left[\frac{i}{\pi} \partial_{x}\theta\, \partial_{\tau} \phi - \mathcal{\hat H}_{0}-\mathcal{\hat H}_{\rm int}- \mathcal{\hat H}_{\rm FQH}-\delta \mathcal{\hat H}\right]  } \nonumber\\
&=&\!\frac{1}{Z} \int \mathcal{D}\phi\,\mathcal{D}\theta  \, \rho^{(1)}_{y_0}(x) \, \left(1 - \int dx   d\tau \,\delta \hat{\mathcal{H} }+ \cdots\right) \nonumber\\
&&\times\,e^{ \int  d\tau \int  dx\, \left[\frac{i}{\pi} \partial_{x}\theta\, \partial_{\tau} \phi - \mathcal{\hat H}_{0}-\mathcal{\hat H}_{\rm int}- \mathcal{\hat H}_{\rm FQH}\right]  } \,\, ,
\label{eq:expansionofthedensitydelta}
\end{eqnarray}
where $Z$ is the partition function, $\mathcal{\hat H}_0$, $\mathcal{\hat H}_{\rm int}$, $\mathcal{\hat H}_{\rm FQH}$ and $\delta\mathcal{\hat H}$ represent the Hamiltonian densities for $\hat H_0$, $\hat H_{\rm int}$, $\hat H_{\rm FQH}$ and $\delta \hat H$, respectively, defined by the usual relation $\hat H=\int dx\, \mathcal{\hat H}$. Higher order in $\delta\mathcal{\hat H}$ are contained in the ellipsis in Eq.~\eqref{eq:expansionofthedensitydelta}. We already noted that the zero order term in $\delta\mathcal{\hat H}$ vanishes, and it corresponds to the FQH ground state.

\begin{figure}[t]
	\centering	
	\includegraphics[width=1\linewidth]{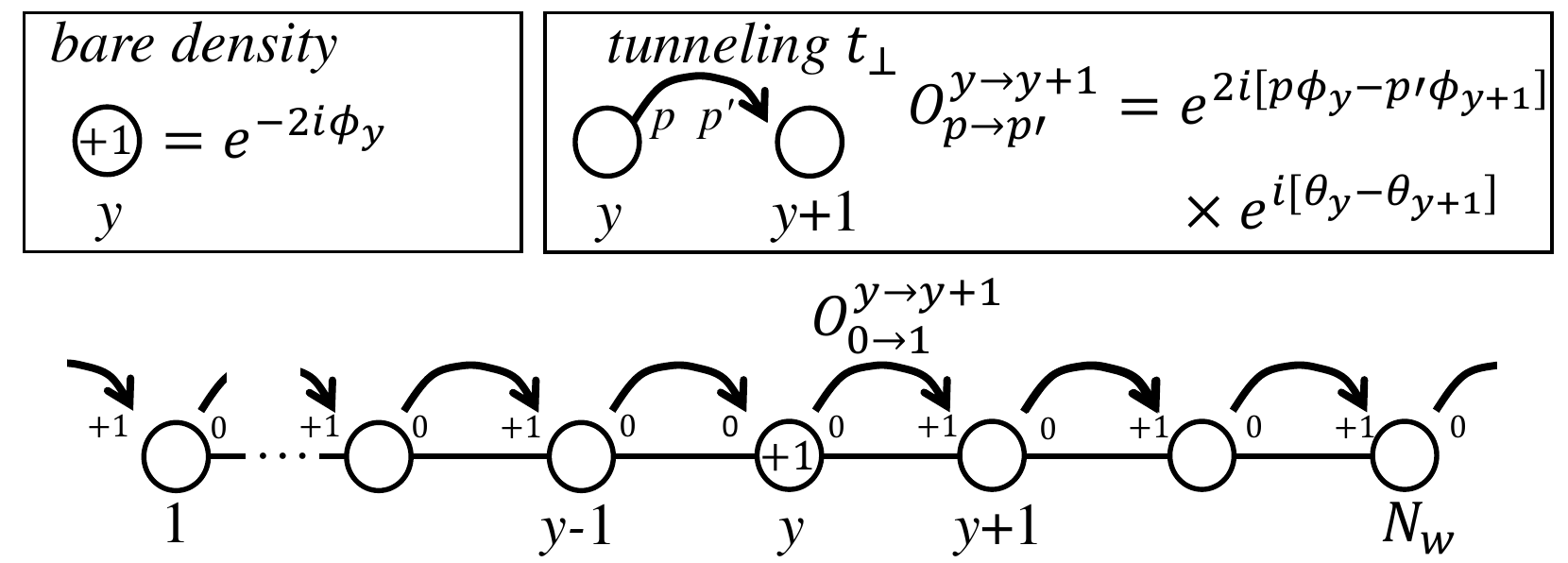}
	\caption{Diagramatic representation of the terms in the sum over $\{p_y,p_y ' \}$ in Eq.~(\ref{eq:densityperturbativeexpansioncorrelationfunction}) contributing to the amplitude $A_{\rm CDW}^{(N_w)}$ in Eq.~(\ref{eq:CDWbasic}).}
	\label{fig:perttheory}
\end{figure}

To obtain the leading term in perturbation theory, which we anticipated to be the loop operator in Eq.~(\ref{eq:productrho}), we keep only $N_w$ link tunneling operators $\mathcal{O}_{p p'}^{y \to y+1}$ in Eq.~\eqref{eq:expansionofthedensitydelta}, which amount to transporting a particle around the cylinder. We furthermore choose $p$ and $p'$ in such a way to generate the operator $e^{-2i \phi_y}$ for each leg $y$ to obtain  Eq.~(\ref{eq:productrho}).
%, except for the leg $y_0$ at which the density operator is evaluated, because the term $e^{-2i\phi_{y_0}}$ is already contained in the expression of the bare density, see Eqs.~\eqref{eq:rhoexp} and~\eqref{eq:expansionofthedensitydelta}.
One obtains
\begin{eqnarray}
&&\left\langle \rho^{(1)}_{y_0}(x)\right\rangle = 2\beta_{1,y_0} t_\perp^{N_w} e^{i\,2 \pi \rho_0 x} \int \prod_{i=1}^{N_w} (dx_i \,d\tau_i) \nonumber\\
&&\hspace{1cm}\times\sum_{\{p_y , p'_y \}}'\,\prod_{y=1}^{N_w}\,C_{p_y,p_y'}^{y,y+1}\, e^{i \Phi x_y}\,e^{-i (p_y - p_y') 2 \pi \rho_0 x_y}\nonumber\\
&&\hspace{1cm}\times\,\left\langle e^{- 2 \,i\, \phi_{y_0}(0,0)}\, \prod_{y=1}^{N_w}\mathcal{O}_{p_y \to  p'_{y}}^{y  \to y+1}(x_y, \tau_y)   \right\rangle ,
\label{eq:densityperturbativeexpansioncorrelationfunction}
\end{eqnarray}
where the Hermitian conjugated terms are not shown, in order to ease the notation. The reason for having the field $\phi_{y_0}$ computed at $x=0$ in the expectation value in Eq.~\eqref{eq:densityperturbativeexpansioncorrelationfunction}, while the oscillating factor outside the integrals is $e^{i\,2 \pi \rho_0 x}$, arises from the fact that, within the perturbative scheme, we assume an infinite translationally-invariant system, and therefore the expectation value is independent of $x$.%We therefore choose $x=0$ for simplicity.

In order to have a nonzero expectation value in Eq.~\eqref{eq:densityperturbativeexpansioncorrelationfunction}, since the computation of the expectation value reduces to that of correlation functions of bosonic fields~\cite{giamarchi2003quantum}, the summation $\sum'_{\{p_y , p'_y \}}$ is restricted to terms that satisfy
\begin{equation}
\label{eq:pppcondition}
p_y-p'_{y-1}=\delta_{y,y_0}-1\,\, ,
\end{equation}
for all $y$. Each term in the sum can be depicted as in Fig.~\ref{fig:perttheory}. In this diagram, we denote the bare density operator $e^{- i 2 \phi_{y_0}}$ by $+1$ on wire $y$, and each arrow corresponds to an operator $\mathcal{O}_{p p'}^{y \to y+1}$.

One now proceeds by transforming the original fields $\{\phi_y\}$ and $\{\theta_y\}$ to $\{\tilde{\phi}_{y\pm\frac{1}{2}}\}$ and $\{\tilde{\theta}_{y\pm\frac{1}{2}}\}$ by means of Eq.~\eqref{transfREV}. Treating the pinned fields $\{\tilde\phi_{y\pm\frac{1}{2}}\}$ as constants, one precisely acquires the factor determining the integer $\kappa$ in Eq.~(\ref{eq:kappa}). The final result of the calculation is the CDW pattern in Eq.~\eqref{eq:CDWbasic}: $\rho(x)~\sim~\rho_0+~A_{\rm CDW}^{(N_w)}~\cos(2 \pi n x -~2 \pi \nu \kappa)$, where, from Eq.~\eqref{eq:densityperturbativeexpansioncorrelationfunction}, the coefficient has the explicit expression
\begin{widetext}
	\begin{equation}
	\label{Eq:ACDW}
	A_{\rm CDW}^{(N_w)} = 2\beta_{1,y_0} t_\perp^{N_w}  \int \prod_{i=1}^{N_w} (dx_i \,d\tau_i) \sum'_{\{p_y , p'_y \}} \left\langle e^{- 2 i \phi_{y_0}(0,0)} \prod_{y=1}^{N_w}\,C_{p_y,p_y'}^{y,y+1}\, \left[ e^{i \Phi x_y} \,e^{-i (p_y - p_y') 2 \pi \rho_0 x_y} \, {\mathcal{O}'}_{p_y \to  p'_{y}}^{y  \to y+1}(x_y, \tau_y)\right] \right\rangle ,
	\end{equation}
\end{widetext}
where the $\{\mathcal{O}'\}$ are obtained from the tunneling link operators in Eq.~\eqref{eq:linktunnelingoperator} by performing the transformation in Eq.~(\ref{transfREV}) and keeping only fluctuating $\tilde{\theta}_{y+\frac{1}{2}}$ fields [the constant $\tilde{\phi}_{y+\frac{1}{2}}$ fields are already inside $\kappa$, see Eq.~\eqref{eq:kappa}].
%From Eq.~\eqref{eq:densityperturbativeexpansioncorrelationfunction}, the CDW structure of the density can be obtained. For the details of the calculation, the reader is referred to Appendix~\ref{appendix:B}. We report here the main results.
The strongly fluctuating fields $\{\tilde\theta_{y\pm\frac{1}{2}}\}$ in the expectation value yield a $(N_w+1)$-point function. It decays exponentially at long distances with a typical correlation length $\xi \sim v/\Delta_{\rm gap}$ determined by the inverse gap $\Delta_{\rm gap}$ opened by the relevant FQH Hamiltonian $\hat H_{\rm FQH}$~\cite{cornfeld2015chiral}.  For details of the calculation of the correlation function, the reader is referred to Appendix~\ref{appendix:B}.

In Sec.~\ref{sec:amplitudeofthecdw} we will present our numerical results on a CDW state for the $N_w=2$ leg ladder. Our goal in the remainder of this section is to use Eq.~(\ref{Eq:ACDW}) to evaluate the amplitude of the CDW in the anisotropic limit $t_\perp \ll t$ and later compare the dependence of $A_{\rm CDW}^{(2)}$ on $t_\perp$ with our numerical results.

\subsection{Amplitude of the CDW for the two-leg ladder}
\label{sec:evaluatingthecdwamplitude}
We now focus on the two-leg ladder ($N_w=2$). As will be discussed in detail in Sec.~\ref{se:2legladder}, we consider the FQH state at filling factor $\nu=1/2$, i.e. $q=1$. In Appendix~\ref{appendix:B}, we compute the three-point correlation function appearing in the CDW amplitude. We obtain
\begin{equation}
\label{xi4}
A_{\rm CDW}^{(2)}= 2\beta_{1,y_0} t_\perp^2 \xi^4 v^{-2} \sum_{p_1, p_1 '}'\,C_{p_1,p_1'}^{1,2}\,C_{p_1'-1,p_1}^{2,1}\, I_{p_1, p_1 '}(\rho_0 \xi) \,\, ,
\end{equation}
where
%\begin{widetext}
%\begin{equation}
%\label{IPP}
%I_{p_1, p_1 '}(\rho_0 \xi)= \int dx_1 d\tau_1 dx_2  d\tau_2  \,e^{2 \pi i \rho_0 \xi x_1 [2 - (p_1 - p_1 ')]}e^{2 \pi i \rho_0 \xi x_2 [2 - (p_1 ' -1 - p_1)]}\,e^{-r_{01}\frac{p_1 + p_1 '}{(2p)^2}} e^{r_{02} \frac{p_1 + p_1 ' -1}{(2p)^2}} e^{-r_{12} \frac{(p_1 + p_1 ')(p_1 + p_1 '-1)}{(2p)^2}} \,\, .
%\end{equation}
%\end{widetext}
\begin{eqnarray}
\label{IPP}
&&\hspace{-0.3cm}I_{p_1, p_1 '}(\rho_0 \xi)= \int dx_1 d\tau_1 dx_2  d\tau_2 \nonumber\\
&&\hspace{0.5cm}\times\,e^{2 \pi i \rho_0 \xi (x_1 [2 - (p_1 - p_1 ')]+x_2 [2 - (p_1 ' -1 - p_1)])}\nonumber\\
&&\hspace{0.5cm}\times\,e^{-r_{01}\frac{p_1 + p_1 '}{(2q)^2}} e^{r_{02} \frac{p_1 + p_1 ' -1}{(2q)^2}} e^{-r_{12} \frac{(p_1 + p_1 ')(p_1 + p_1 '-1)}{(2q)^2}}.
\end{eqnarray}
Here, $r_{ij}=\sqrt{(x_i - x_j)^2+v^2(\tau_i - \tau_j )^2}$ and $r_{i}=\sqrt{x_i^2+v^2 \tau_i  ^2}$, where $x_i$ and $v \tau_i$ are dimensionless variables obtained by  $x_i \to  x_i / \xi$, and similarly for $\tau_i$. In Eq.~\eqref{IPP}, the second line contains oscillating factors controlled by the dimensionless variable $\kappa=\rho_0 \xi$. The third line contains exponential factors which separately either decay or diverge,  but overall the integrand decays exponentially as any of the coordinates is sent to infinity.

Our focus now is to extract from Eq.~\eqref{IPP} the $t_\perp$ dependence of the CDW amplitude. Apart from the explicit $t_\perp^{2}$ dependence of $A_{\rm CDW}^{(2)}$, the correlation length $\xi = v/\Delta_{\rm gap}$ also depends on $t_\perp$ through the energy gap $\Delta_{\rm gap} \sim t \left(t_\perp/t \right)^{1/(2-X_{\rm FQH})}$ with $0<X_{\rm FQH}<2$ being the scaling dimension of the relevant FQH operator~\cite{petrescu2015chiral,cornfeld2015chiral}. We thus need to consider the dependence of the integral $I(\rho_0 \xi)$ on  $\kappa$. We have two limits: for $\kappa \ll 1$ the oscillating factors in the integral can be neglected and the integral acquires a finite value, which is just a dimensionless number of order unity.
The sum over $p_1, p_1'$, including also the non-universal coefficients $\{C_{p,p'}\}$, is expected to be finite. Up to this overall non-universal coefficient, we have $A_{\rm CDW}^{(2)} \sim  t_\perp^2\xi^4/v^2 $ for $\kappa \to 0$. This limit of short correlation length, however, corresponds to large $t_\perp$ and hence the wire construction approach which is perturbative in $t_\perp$ is not immediately valid.

Instead, consider the opposite limit  $\kappa \gg 1$, i.e. $\rho_0 \xi \gg~1$, corresponding to small $t_\perp$ and to a long correlation length, where the wire construction approach is controlled. The oscillating factors in $I_{p_1, p_1 '}(\kappa)$ lead to a suppression of the integral in powers of $1/\kappa$. We estimate this limit in Appendix~\ref{appendix:evaluationoftheintegral} and find that $I_{p_1, p_1 '}(\kappa) \propto \kappa^{-5}$. In this limit, the CDW  behaves as
\be
\label{cdw2}
A_{\rm CDW}^{(2)} \sim  \left( \frac{t_\perp}{t} \right)^2 \frac{\Delta_{\rm gap}}{t} \sim   \left( \frac{t_\perp}{t} \right)^{2+ \frac{1}{2-X_{\rm FQH}}} \,\, .
\ee
This dependence on $t_\perp$ for the two-leg ladder case will be compared with our numerical results in the next section. Generalizing for $N_w$ wires in Appendix~\ref{appendix:evaluationoftheintegral}, we find
\be
\label{cdwnw}
A_{\rm CDW}^{(N_w)} \sim \left( \frac{t_\perp}{t} \right)^{N_w} \frac{\Delta_{\rm gap}}{t} \sim \left( \frac{t_\perp}{t} \right)^{N_w+ \frac{1}{2-X_{\rm FQH}}}.
\ee
We can see that as expected the CDW amplitude decays exponentially with the number of wires, and vanishes in the topological 2D limit of $N_w\rightarrow\infty$. This can be written as $e^{- N_w/ N^*}$ with transverse correlation length $\xi_\perp \equiv N^* = 1/\log(t/t_\perp)$.
	In the anisotropic limit of small $t_\perp$, the transverse correlation length becomes very small. This means that even a thin cylinder can be in the topological regime, see Fig.~\ref{fig:1}. This of course comes with a trade-off, since in this limit the energy gap becomes small too, and so the longitudinal correlation length becomes large, requiring long systems.

To summarize this section, in the thin cylinder limit there is a CDW, whose
phase shift measures the eigenvalues of the Wilson loop operator. In our
calculation, we assumed that the system is in a specific eigenstate. On an
infinite homogeneous system or on a torus, these $2q$ states are degenerate.
For OBC in the \rev{real dimension $x$ (or $j$ on the lattice)}, the physics at the boundaries can break the degeneracy and the system
chooses one state, as we will see in the next section. Alternatively, consider
an infinite cylinder with an extra potential $\mu_{j}$ at site $j$ on wire $y$.
In its presence there is a splitting of the energies of the $2q $ CDW states
such that $E = E(\kappa) =\mu_{j}\, A_{\rm{CDW}}^{(N_w)}\,\cos(2 \pi n j - 2
\pi \nu \kappa) + {\rm const}$. This means, that the exponentially small
amplitude $A_{\rm{CDW}}^{(N_w)}$, becomes also the coefficient of a term in the
Hamiltonian, that contains the nonlocal Wilson loops. Such local potentials
will be utilized in the next chapter to control the ground states.

\section{Precursors of topology on the bosonic two-leg ladder}
\label{se:2legladder}
In this section, we explore the connection between the local order parameter regime and the topological regime in the extremely (quasi 1D) thin limit: a two-leg ladder ($N_w=2$). We present a simple way to measure fractional charge excitations in the $\nu=1/2$ Laughlin-like state of hard-core bosons in the two-leg flux ladder. A possible way to create and measure fractional excitations with charge $1/2$ is to create interfaces between the two different CDW ground states~\cite{PhysRevB.28.2264,PhysRevB.30.1069,PhysRevB.32.2617,RevModPhys.91.015005} that are expected to arise when a full gap in the low-energy spectrum of the Laughlin-like state is induced, i.e., by closing the FQH state on a thin torus. By resorting to an extensive numerical analysis by using a MPS-based algorithm~\cite{SCHOLLWOCK201196}, we create such domains walls in our system and measure $1/2$ fractional charge excitations. We argue that such fractional charge excitations are connected to Laughlin quasiparticles in the topological regime in the limit of large $N_w$ or small $t_\perp / t$.

\subsection{Model for fully-gapped Laughlin-like state}
In this section, we consider the thin \rev{cylinder} limit of the two-leg flux ladder~\cite{grusdt2014realization,petrescu2017precursor}. In order to achieve such limit, we consider the Hamiltonian in Eq.~\eqref{eq:hamiltonianfermionctwolegladder} for $N_w=2$ with the inclusion of a space-dependent transverse hopping parameter $t_\perp\rightarrow t_\perp(j)=t_\perp+t_{\alpha}\,e^{-i\alpha j}$, with real $t_\perp$ and $t_\alpha$:
\begin{eqnarray}
\hat H&=&-t\sum_{j=1}^{L-1}\,\sum_{y=1,2} \hat b^\dag_{j,y}\hat b_{j+1,y} + \sum_{j=1}^{L}t_\perp(j)\,
\hat b^\dag_{j,2}\hat b_{j,1}\,e^{i\Phi j}\nonumber\\
\nonumber\\
&&+\rev{V_\perp}\sum_{j=1}^{L}\hat n_{j,1}\hat n_{j,2} +{\rm H.c.}\,\, ,
\label{eq:hamiltonianlaughinlikestate}
\end{eqnarray}
where, in this case, $\hat b_{j,y}$ ($\hat b^\dag_{j,y}$) represents the annihilation (creation) operator of a hard-core boson on site $j$ and leg $y=1,2$, $V_\perp$ represents an inter-leg density-density interaction, where $\hat n_{j,y}=\hat b^\dag_{j,y}\hat b_{j,y}$ is the particle density operator on site $j$ and leg $y$. In order to induce a full gap in the low-energy spectrum, as we explain below, we choose $\alpha=8\pi\rho_0$.

The fact that the Hamiltonian in Eq.~\eqref{eq:hamiltonianlaughinlikestate} realizes the thin torus limit on a two-leg ladder can be understood by expanding the inter-chain hopping operators following the field theory approach in Sec.~\ref{se:1D2D}:
\begin{eqnarray}
\label{eq:Oppp1}
\hat{H}_\perp &=&  \int dx\,t_\perp(x)\,\Psi_{2}^\dagger(x)\Psi_{1}(x)\,e^{i \Phi x} +{\rm H.c.} \nonumber \\
&=& \int dx  \left[t_\perp\,e^{i \Phi x}+t_\alpha\,e^{i (\Phi-\alpha) x}\right] \nonumber\\
&&\times\,\sum_{p  , p'} e^{ -i (p-p') 2 \pi \rho_0 x} \mathcal{O}_{p p'}^{1 \to 2}+{\rm H.c.} \,\, .
\end{eqnarray}
For a spatially uniform $t_\perp(x)$, i.e. $t_\alpha=0$, only the FQH operator $\mathcal{O}_{pp'}^{1 \to 2}$ with $p ' = -p$ where $p\equiv q={(2\nu)}^{-1}=1$, which is $\mathcal{O}_{p,p'}^{1 \to 2}\sim e^{i\,2\tilde\phi_{1/2}}$ [Eq.~\eqref{eq:phitheta2leg}], becomes non-oscillating for filling factor $\nu = 2\pi\rho_0/\Phi=1/2$, see Eqs.~\eqref{eq:Oppp} and~\eqref{eq:Hperp}, resulting in the gapping of the link field $\tilde\phi_{\frac{1}{2}}$. In this case, the model in Eq.~\eqref{eq:hamiltonianlaughinlikestate} is predicted to display the one-dimensional analog of the Laughlin state (the \emph{Laughlin-like state}) when $\nu=1/2$~\cite{petrescu2013bosonic,cornfeld2015chiral}. Such state has been detected in the flux ladder by observing the universal two-cusp behavior of the chiral current and entanglement-related observables (central charge)~\cite{strinati2017laughlin}, signalling the Lifshitz commensurate-incommensurate transition~\cite{giamarchi2003quantum} from a standard gapless phase, to a helical partially-gapped phase, when the commensurability condition $\Phi=4\pi \rho_0$ is met. However, a direct measurement of excitations with fractional charge $\nu=1/2$ has not been provided yet.

The gapping of the second link field $\tilde{\phi}_{-\frac{1}{2}}$ is achieved by taking $\mathcal{O}_{pp'}^{2 \to 1}$ with $p' = -p =  (2 \nu)^{-1}=1$, i.e., $\mathcal{O}_{p,p'}^{2 \to 1}\sim e^{i\,2\tilde\phi_{-1/2}}$ in the expansion in Eq.~\eqref{eq:Oppp1}. This latter term, which is always oscillating for $t_\alpha=0$ and therefore irrelevant, can be made non-oscillating at the $\nu=1/2$ Laughlin-like state instability by the presence of the additional oscillating phase $e^{i(\Phi-\alpha)x}$ when $t_\alpha\neq0$ in Eq.~\eqref{eq:Oppp1}, by choosing $\alpha=8\pi\rho_0=2 \Phi$. A similar mechanism was used in Ref.~\cite{oreg2014fractional}. In the following, we choose $t_\alpha=t_\perp$. In this case, around the commensurability condition $\Phi=4\pi\rho_0$, both fields $\tilde\phi_{\pm\frac{1}{2}}$ are gapped, and the fully-gapped $\nu=1/2$ Laughlin-like state is achieved.

\subsection{Numerical results}
In this section, we discuss our numerical results. In order to obtain the ground state of the Hamiltonian in Eq.~\eqref{eq:hamiltonianlaughinlikestate}, we use the MPS-based algorithm following the same scheme as in Ref.~\cite{strinati2017laughlin}. We recall below the procedure for the sake of completeness. We consider OBC along the $j$ direction, and we initialize the system in a random MPS state with initial bond link $D_{\rm in}=150$, and then perform an imaginary-time evolution up to time $100\,t^{-1}$ with maximum bond link $D_{\rm im,max}=200$. The ground state of the system is found after a local variational search in the MPS space sweeping the chain until convergence is reached, i.e., until the ground-state energy approaches a constant value. In our simulations we fix the number of lattice sites $L$ along the $j$ direction, which corresponds to the number of plaquettes, the gauge flux $\Phi$, the transverse tunnelling amplitude $t_\perp$, the on-site interaction strength $V_\perp$, and the maximum value of the bond link in the variational procedure, $D_{\rm max}$, which we use to approximate the final MPS ground state. Also, since the total number of particles $N=\langle\sum_j\,\sum_m\,\hat n_{j,m}\rangle$ is a conserved quantity, %[i.e., the Hamiltonian in Eq.~\eqref{eq:hamiltonianlaughinlikestate} commutes with the total density operator $\sum_j\,\sum_m\,\hat n_{j,m}$],
in our numerical simulations we work at fixed $N$. \rev{The hard-core-boson constraint is implemented by limiting the dimension of the local Hilbert space to $4$, on each rung (see Fig.~\ref{fig:cylinderscheme})}. As argued in Ref.~\cite{strinati2017laughlin}, the choice of the value of $D_{\rm max}$ plays a crucial role in the computation of the entanglement entropy, but has a less drastic effect on the computation of local and two-point correlators, such as densities or chiral currents. In what follows, if not explicit, we use $t$ as a reference energy scale, and set $\hbar=1$.

\subsection{Controlling the ground state using external local chemical potentials}
\label{sec:controllingthegroundstatewithexternalchemicalpotentials}
\begin{figure}[t]
	\centering
	\includegraphics[width=7.5cm]{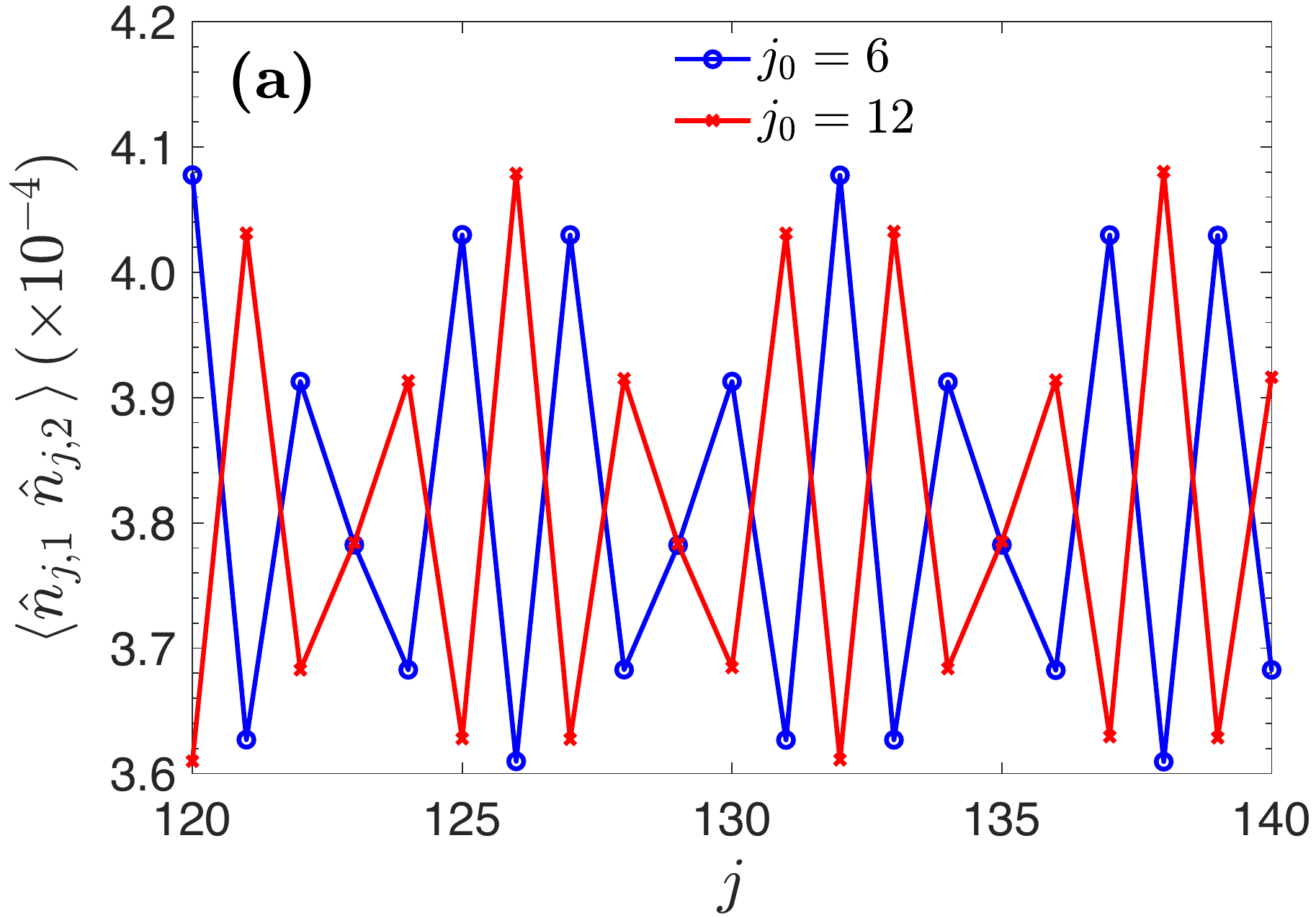}
	\includegraphics[width=7.5cm]{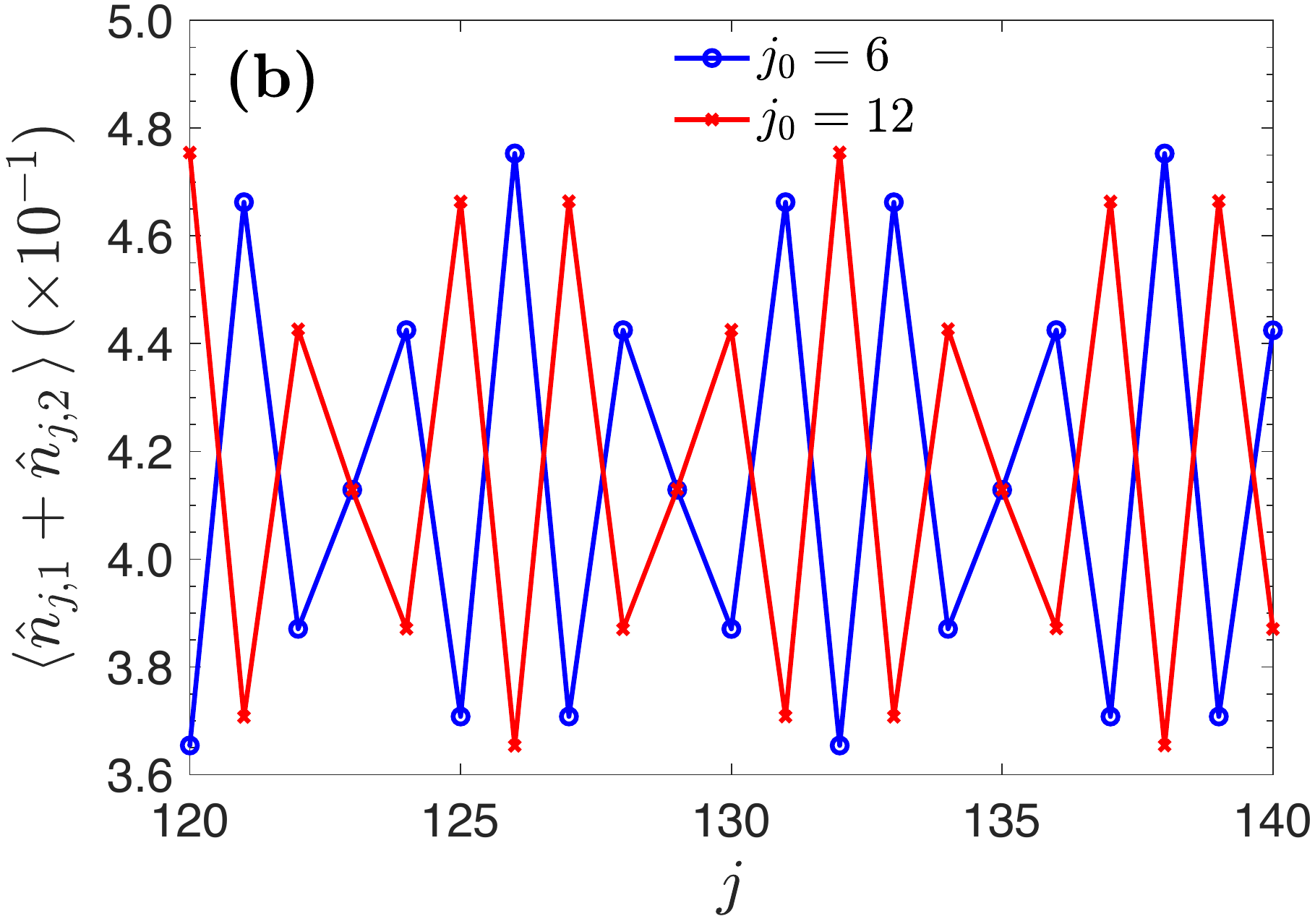}
	\caption{Data series for \textbf{(a)} $\langle\hat n_{j,1}\hat n_{j,2}\rangle$ and \textbf{(b)} $\sum_{y=1,2}\langle\hat n_{j,y}\rangle$, with $j$ around $j=130$, for a simulation with $L=240$, $N=~100$ (i.e., $n=N/L=5/12$), $t_\perp=10^{-1}\,t$, $V_\perp=30\,t$ and $\Phi/\pi\simeq0.832$. The expectation values are computed on the $|\Psi_{\rm CDW_1}\rangle$ ground state (blue data) or on the $|\Psi_{\rm CDW_2}\rangle$ ground state (red data), see text. The two CDWs, \rev{which appear with spatial period $\lambda=12$ on the lattice}, are numerically obtained by using a boundary chemical potential $\mu=-0.4\,t$, on the left chain end only, on two sites: $j=6,18$ for the blue data (i.e., $j_0=6$ and $r=0,1$), and $j=12,24$ for the red data (i.e., $j_0=12$ and $r=0,1$).}
	\label{fig:datal240shargedensitywaves}
\end{figure}
 We first present numerical data for $L=240$, $N=~100$ (i.e., $n=N/L=5/12$), $t_\perp=10^{-1}\,t$, $V_\perp=30\,t$ and set $\Phi/\pi\simeq0.832 $ in order to drive the system to the commensurate Laughlin-like state $\Phi/ \pi = 4\rho_0=2n$. With such large value of $L$, since we are not interested in measuring entanglement-related observables, we use $D_{\rm max}=200$ in order to reduce the numerical complexity of the problem. We measure both the total particle densities, $\sum_y\hat n_{j,y}$, and the local product of the two densities, $\hat n_{j,1}\hat n_{j,2}$. As we see from Eq.~(\ref{eq:productrho}), for the $\nu=1/2$ Laughlin-like state, the two ground states consist of two CDWs with equal spatial period $\lambda\propto1/(2\rho_0)$ (which is numerically obtained from the CDW data, and it is of $\lambda=12$ sites in our case, and sites within a period identify a \emph{unit cell}) related to the particle density $n=2\rho_0$, but one is shifted by 6 sites (i.e., half unit cell, $\lambda/2$) with respect to the other one: we call these two ground states $|\Psi_{\rm CDW_1}\rangle$ and $|\Psi_{\rm CDW_2}\rangle$.

Because of OBC along the $j$ direction, these two states are in fact not exactly degenerate for a finite system. In order to select one of the two admitted CDW patterns, and therefore control the ground state at which the algorithm converges, we add to the Hamiltonian in Eq.~\eqref{eq:hamiltonianlaughinlikestate} a local chemical potential of the form $\hat H_{\rm loc}=\sum_j\,\sum_{y=1,2}\mu_j$, where $\mu_j=\mu<0$ for $j=j_0+r\lambda$, for some integer $r$ and $j_0$, whereas $\mu_j=0$ otherwise. Specifically, we put a nonzero chemical potential only on a few sites close to the boundaries of the system. The selection of the ground state is therefore understood: if for a given $j_0$ the numerical algorithm converges to $|\Psi_{\rm CDW_1}\rangle$, the convergence to the other ground state $|\Psi_{\rm CDW_2}\rangle$ is enforced by using for example $j_0\rightarrow j_0+\lambda/2$.

The data of the simulations are shown in Fig.~\ref{fig:datal240shargedensitywaves}. In particular, we show in panel \textbf{(a)} the data series for $\langle\hat n_{j,1}\hat n_{j,2}\rangle$, and $\sum_{y=1,2}\langle\hat n_{j,y}\rangle$ in panel \textbf{(b)}, using the parameters listed in the caption. The fact that $\langle\hat n_{j,1}\hat n_{j,2}\rangle\ll\sum_{y=1,2}\langle\hat n_{j,y}\rangle$ is a consequence of the large value of $V_\perp$ that we use~\cite{strinati2017laughlin}. The data are shown as a function of the site label $j$, and we focus only on some bulk sites around site $j=130$ for clarity. The expectation values are computed on the $|\Psi_{\rm CDW_1}\rangle$ ground state (blue data) or on the $|\Psi_{\rm CDW_2}\rangle$ ground state (red data). The two CDWs are numerically obtained by using a boundary chemical potential $\mu=-0.4\,t$, on the left chain end only, on two sites: $j=6,18$ for the blue data, and $j=12,24$ for the red data. We therefore obtain two perfect CDWs, sufficiently far away from the chain ends, with period $\lambda=12$ and that are shifted by $\lambda/2=6$ sites, as predicted by the bosonization arguments presented in Sec.~\ref{se:1D2D}.

\begin{figure}[t]
	\centering
	\includegraphics[width=8cm]{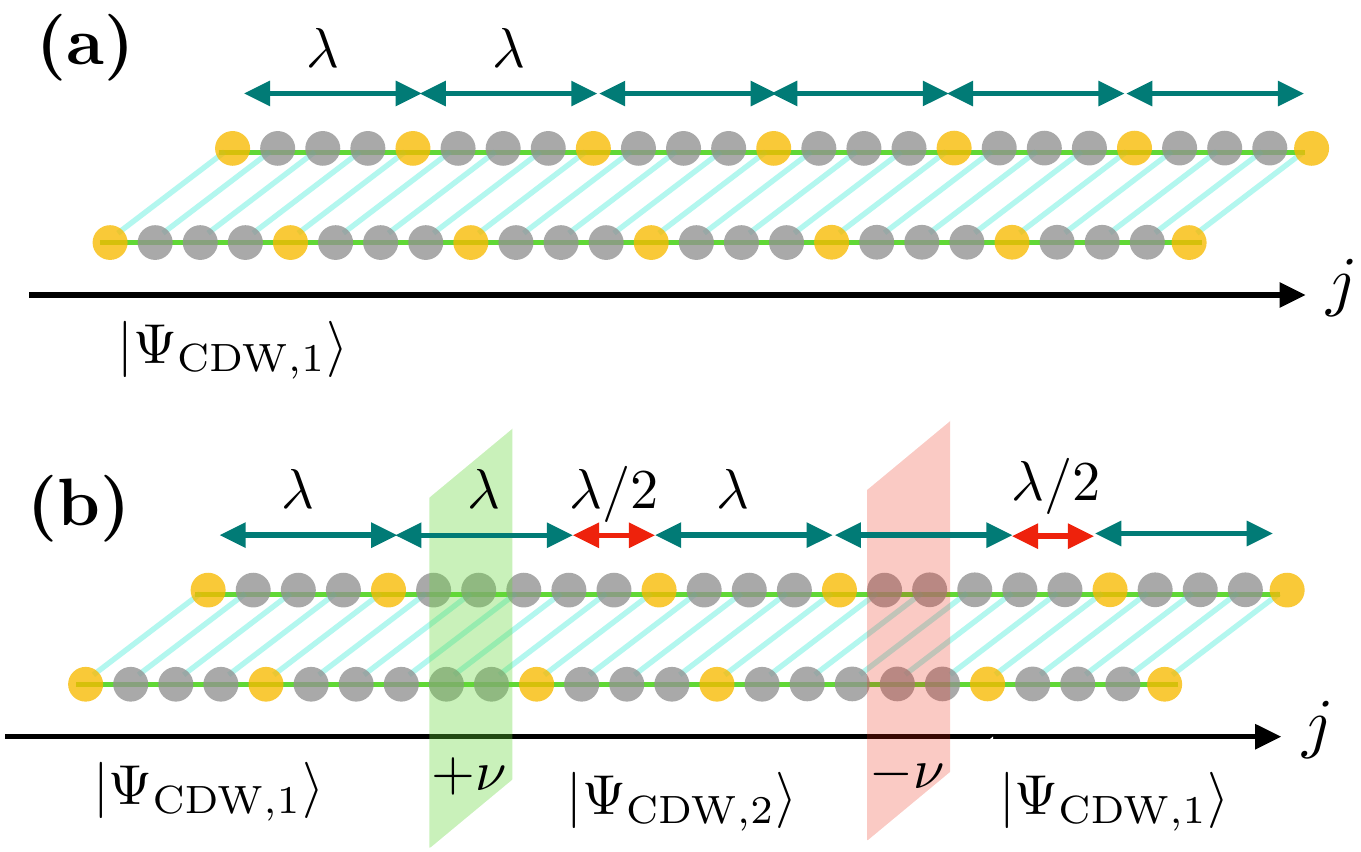}
	\caption{Schematic representation of the formation of the domain wall on the two-leg ladder. Grey dots represents the sites of the chains along the longitudinal direction $j$, cyan lines are the $t$ and $t_\perp$ links, and yellow sites are the sites at which local chemical potential is applied. {\textbf{(a)}} One of the two CDW patterns in Eq.~\eqref{eq:CDWbasic}, e.g. $|\Psi_{\rm CDW,1}\rangle$, can be chosen by applying the local potential only on sites with a relative distance equal to the size of the unit cell $\lambda$. {\textbf{(b)}} From the configuration as in panel \textbf{(a)}, the local chemical potential on the central region of the ladder is displaced by $\lambda/2$ with respect to the previous configuration, therefore enforcing the other CDW pattern, $|\Psi_{\rm CDW,2}\rangle$, whereas on the two outer regions $|\Psi_{\rm CDW,1}\rangle$ is chosen as before. Domain walls (green and red plane), carrying opposite fractional charge $\pm\nu$, are found at the interfaces between the different ground states. \rev{Notice that the value of $\lambda=4$ used in the figure is chosen merely for graphical purposes, and it does not reflect the actual value $\lambda=12$ used in the numerical simulation (see text).}}
	\label{fig:domainwallformationscheme}
\end{figure}

\begin{figure*}[t]
	\centering
	\includegraphics[width=17.9cm]{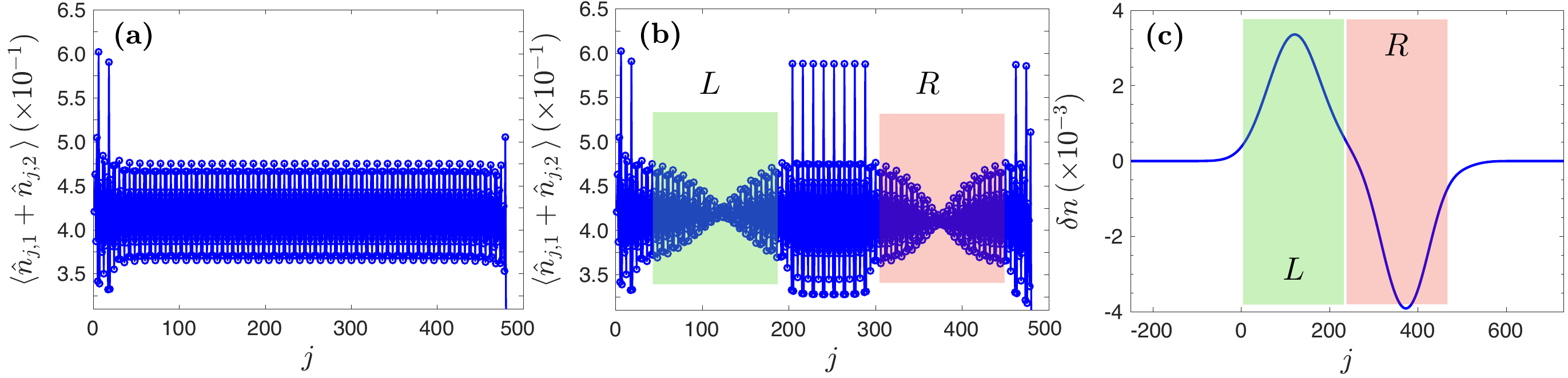}
	\caption{Numerical simulation of the domain wall formation and measurement of the fractional charge. We use $L=480$, $N=200$, $t_\perp=10^{-1}\,t$, $V_\perp=30\,t$, $\Phi/\pi\simeq0.832$ and $D_{\rm max}=200$. We show \textbf{(a)} the total density $\sum_{y=1,2}\langle\hat n_{j,y}\rangle$ measured on $|\Psi_{\rm CDW_1}\rangle$ (without domain walls) and \textbf{(b)} measured with two domain walls, as explained in Fig.~\ref{fig:domainwallformationscheme}. We highlight in the panel the two domain walls, left (L) and right (R) by the green and red shaded area, respectively. The position of the high peaks in the CDW patterns correspond to the sites where $\mu^{(L)}_j$, $\mu^{(B)}_j$ and $\mu^{(R)}_j$ are applied, see text. \textbf{(c)} Local density variation $\delta n_j$ [Eq.~\eqref{eq:deltandomainwall}] computed by subtracting the smeared density in \textbf{(a)} from that in panel \textbf{(b)}. The excess and depletion of particle density in the vicinity of the domain walls appear. The smearing procedure is done by using the Gaussian kernel in Eq.~\eqref{eq:kerneldiscretesum} with $\sigma=3\lambda=36$.}
	\label{fig:domainwallformationscheme2}
\end{figure*}

\subsection{Two-domain-wall structure and fractional charge measurement}
The data in Fig.~\ref{fig:datal240shargedensitywaves} suggest that we can enforce a given CDW ground state by applying a local chemical potential on some sites of the chain. By extending such argument, we can selectively enforce different CDWs in different sub-regions of the system by combining different local chemical potentials on different parts of the chains, therefore creating domain walls, i.e., interfaces between the two different CDW patterns, which host fractional $1/2$ charge excitation. Because of the conservation of the total number of particles, in order to ensure that the overall density is conserved, the minimal configuration consists of two domain walls that carry fractional charge $\pm1/2$ and $\mp1/2$, respectively.

This is done as sketched in Fig.~\ref{fig:domainwallformationscheme}: the local chemical potential is applied at some sites close to the edges and close to the sites at mid-chain, i.e., the bulk sites. At the edges, one uses $\mu^{(R,L)}_j=j_0+r_{R,L}\lambda$ for some integer $r_R$ and $r_L$ such that $\mu_j=\mu=-0.4\,t$ only for $j$ in the vicinity of the right or left edge, respectively, and zero otherwise, whereas the bulk chemical potential is shifted by $\lambda/2$ sites with respect to the edge chemical potentials, i.e., $\mu^{(B)}_j=j_0+\lambda/2+r_{B}\lambda$, for $r_B$ such that $\mu_j=\mu=-0.4\,t$ only for $j$ in the vicinity of the sites at mid-chain, whereas it is zero otherwise. Since the domain walls are expected to have a finite correlation length $\xi$, we need to use a sufficiently large value of $L$ that allows us to clearly resolve two domain walls, while keeping a reasonable numerical complexity, which is also granted by using a not too large value of the bond link $D_{\rm max}$. For this simulation, we therefore use $L=480$, $N=200$, $t_\perp=10^{-1}\,t$, $V_\perp=30\,t$, $\Phi/\pi\simeq0.832$ and $D_{\rm max}=200$.

In order to measure the excess or depletion of charge (density) at each domain wall, we resort to the computation of the total particle density, $\sum_{y=1,2}\langle\hat n_{j,y}\rangle$. By using Eqs.~\eqref{eq:rhoexp} and~\eqref{basistransformation}, the total excess charge between two points $x_1$ and $x_2>x_1$ such that the domain wall is found in between these two points is
\begin{eqnarray}
\label{eq:avergagedensityfillingfactor2}
Q &=&\int_{x_1}^{x_2} dx \,\sum_{y=1,2}  (\rho_y - \rho_0) =-\frac{1}{\pi}\sum_{y=1,2}\int_{x_1}^{x_2} dx\,\partial_x {\phi}_{y} \nonumber\\
&=& \nu \Delta \kappa \,\, .
\end{eqnarray}
This means that interfaces between CDWs localize fractional charges $\nu$.

The numerical procedure that we follow in order to measure the fractional charge $\nu$ is the following: (i) we simulate the Hamiltonian Eq.~(\ref{eq:hamiltonianlaughinlikestate}) with only an edge chemical potential [e.g., $\mu^{(L)}_j$] and obtain the pattern of the CDW without domain walls, i.e., $n_{j,1}=\sum_{y=1,2}\langle\Psi_{\rm CDW_1}|\hat n_{j,y}|\Psi_{\rm CDW_1}\rangle$ [Fig.~\ref{fig:domainwallformationscheme2}, panel \textbf{(a)}]. Then (ii) we put three local chemical potentials, $\mu^{(L)}_j$, $\mu^{(B)}_j$ and $\mu^{(R)}_j$ on the left edge, bulk, and right edge sites, respectively, such that the edge chemical potentials locally enforce the $|\Psi_{\rm CDW_1}\rangle$ pattern at the left and right chain ends, whereas the bulk chemical potential is shifted by $\lambda/2=6$ sites with respect to the edge ones in order to locally enforce the $|\Psi_{\rm CDW_2}\rangle$ pattern, creating two domain walls where the two patterns merge (recall Fig.~\ref{fig:domainwallformationscheme}). We call $|\Psi_{\rm DW}\rangle$ the resulting ground state. The resulting pattern of the total particle density, $n_{j,{\rm DW}}=\sum_{y=1,2}\langle\Psi_{\rm DW}|\hat n_{j,y}|\Psi_{\rm DW}\rangle$, is shown in Fig.~\ref{fig:domainwallformationscheme2}, panel \textbf{(b)}.

The fractional charge is measured by first computing the macroscopic (\emph{smeared}) densities $n_j \to n_{s , j} = \sum_h K_{j,h}  n_h$ where $K_{j,h} \propto e^{- (j-h)^2/(2 \sigma^2)}$ [Eq.~\eqref{eq:smeafeddensity2}] from $n_{j,1}$ and $n_{j,{\rm DW}}$, by using the Gaussian kernel in Eq.~\eqref{eq:kerneldiscretesum}, with a given variance $\sigma$. In order to ensure the correct normalization of the kernel, and therefore the conservation of the number of particle after the smearing procedure, we use $L_0=250$ auxiliary ghost sites on both chain ends in the computation of the macroscopic densities, see Appendix~\ref{app:smeareddensity} for more details. We then define
\begin{equation}
\begin{array}{l}
\displaystyle{n_{s,j,1}=\sum_{h=-L_0}^{L+L_0}K_{j,h}\,n_{h,1}}\\
\displaystyle{n_{s,j,{\rm DW}}=\sum_{h=-L_0}^{L+L_0}K_{j,h}\,n_{h,{\rm DW}}}
\end{array} \,\, ,
\label{eq:smeareddesitiesdw}
\end{equation}
where $j=-L_0,\dots,L+L_0$ and sites for $j<0$ and $j>L=480$ should be intended as ghost sites. The precise $j$ dependence of the smeared quantities in Eq.~\eqref{eq:smeareddesitiesdw} depends, in this case, on the choice of the width $\sigma$ of the kernel. In order not to be sensitive to variations of the density on length scales of the order of the lattice constant $a$, while resolving single domain walls, we see that $\sigma$ should be chosen such that $a<\sigma<\xi$. Specifically, $\sigma$ is chosen to be of the order of one unit cell. From the macroscopic quantities in Eq.~\eqref{eq:smeareddesitiesdw}, we define
\begin{equation}
\delta n_j=n_{s,j,{\rm DW}}-n_{s,j,1} \,\, .
\label{eq:deltandomainwall}
\end{equation}
The behavior of $\delta n_j$ is shown in Fig.~\ref{fig:domainwallformationscheme2}, panel \textbf{(c)}, in particular for $\sigma=3\lambda=36$. The site $j_S$ that separates the two regions of the two domain walls is estimated by the condition $\delta n_{j_S}=0$. From the data in Fig.~\ref{fig:domainwallformationscheme2}, we therefore compute the charge excess or depletion at the two domain walls as
\begin{equation}
Q_L=\sum_{j=-L_0}^{j_S}\delta n_j \qquad Q_R=\sum_{j=j_S}^{L+L_0}\delta n_j \,\, .
\label{eq:expressionforqlandqr}
\end{equation}

\begin{figure}
\centering
\includegraphics[width=7.5cm]{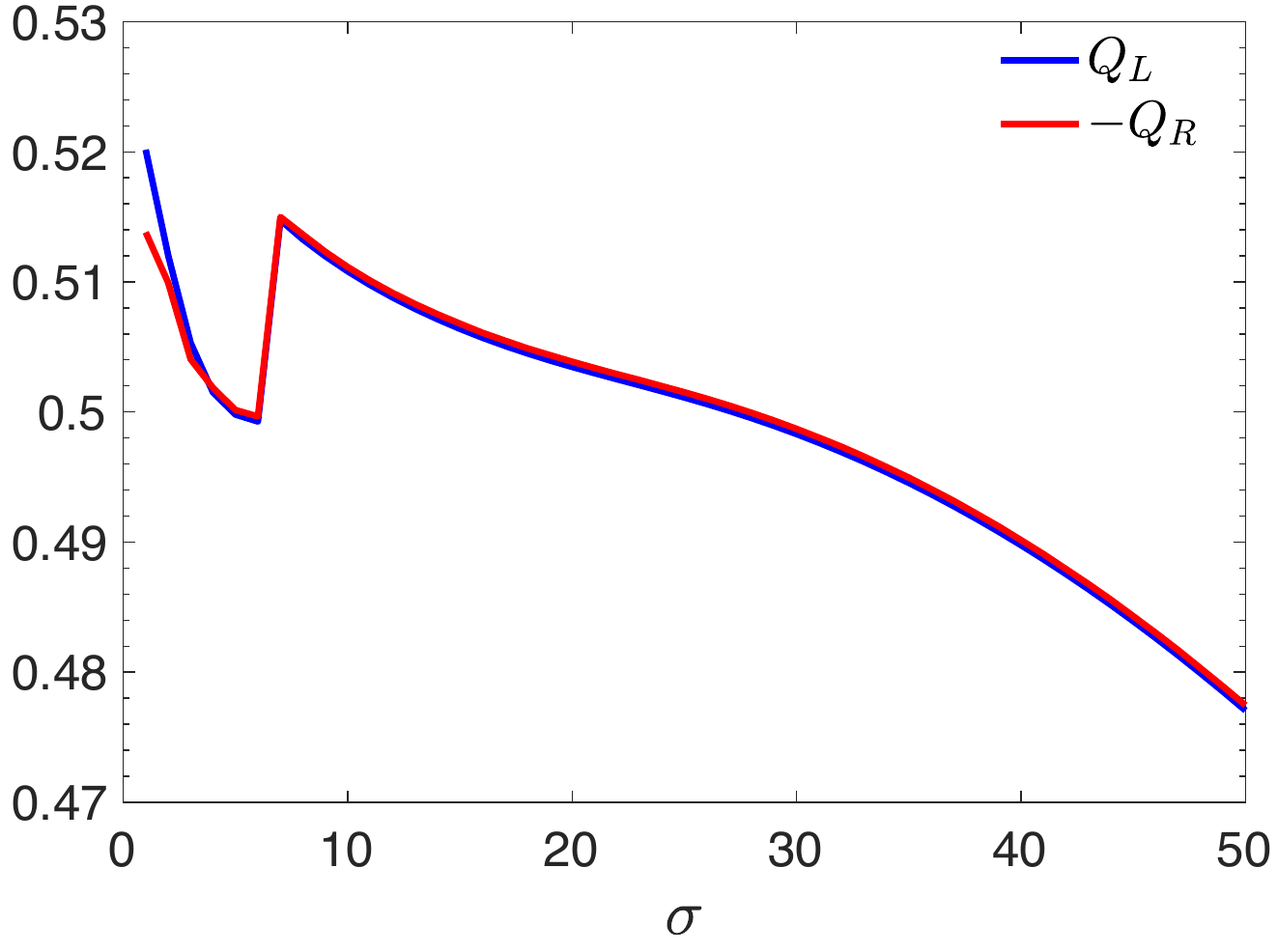}
\caption{Value of the fractional charge for the left (blue data) and right (red data) domain wall in Fig.~\ref{fig:domainwallformationscheme2}, computed using Eq.~\eqref{eq:smeareddesitiesdw} and~\eqref{eq:expressionforqlandqr}, as a function of $\sigma$ in Eq.~\eqref{eq:kerneldiscretesum} in units of the lattice constant $a=1$. We see that, for $\sigma$ of the order of the lattice spacing $a=1$, the computed charge fluctuates, and then it becomes a monotonous decreasing function of $\sigma$, and $Q_L\simeq-Q_R$ is correctly found. For $\lambda\lesssim\sigma\lesssim3\lambda$, the computed charges agree with the expected value $Q_{L,R}=\pm1/2$.}
\label{fig:sigmaanalysis}
\end{figure}

The numerical computation of the excess of density at the domain walls is reported in Fig.~\ref{fig:sigmaanalysis}. We show $Q_L$ (blue data) and \rev{$-Q_R$} (red data) as in Eq.~\eqref{eq:expressionforqlandqr} using different values of $\sigma$ in Eq.~\eqref{eq:kerneldiscretesum}, for the domain walls in Fig.~\ref{fig:domainwallformationscheme2}. We see that, for $\sigma$ of the order of the lattice spacing $a=1$, the computed charge fluctuates, and then it becomes a monotonous decreasing function of $\sigma$, with $Q_L\simeq-Q_R$. For $\lambda\lesssim\sigma\lesssim3\lambda$, the computed charges are in good agreement with the expected value $Q_{L,R}=\pm1/2$\rev{.}

We stress that the correct measurement of $\nu$ is provided only for $a<\sigma<\xi$. Indeed, for $\sigma\sim a$, the microscopic fluctuations are resolved and the form of $\delta n_j$ in Fig.~\ref{fig:domainwallformationscheme2}, panels \textbf{(c)}, will not be a smooth function of $j$, whereas for $\sigma$ sufficiently larger than $\xi$, the procedure would also include sites that are not part of the domain walls, and such inclusion will prevent us from clearly resolving the single domain walls, as required by Eq.~\eqref{eq:deltandomainwall}. This is evident in the $\sigma\rightarrow\infty$ limit: Indeed, we expect $n_{s,j,{\rm DW}}=~n_{s,j,1}$ in the very large $\sigma$ limit, since the results tend to be independent of $j$ and equal to $N$.

\begin{figure}[t]
\centering
\includegraphics[width=7.5cm]{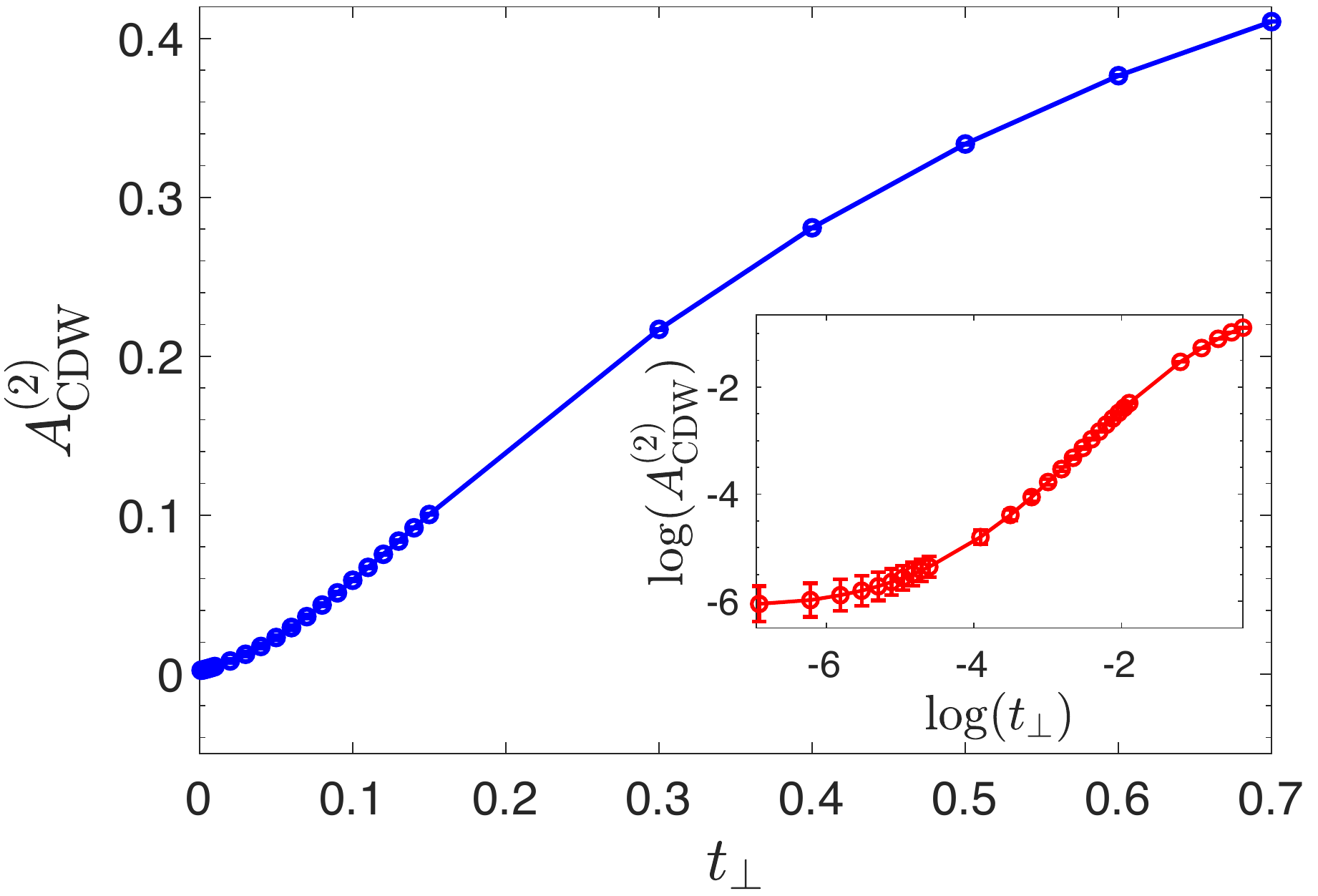}
\caption{Numerical data for $A_{\rm CDW}^{(2)}\equiv\max_{j\in I}\langle\sum_y\hat n_{j,y}\rangle-N/L$, where $I=[\Delta L:L-\Delta L]$ is a subregion of the chain to which the sites close to the ends have been \rev{removed}, in order to avoid boundary effects. The data are obtained by simulating the Hamiltonian in Eq.~\eqref{eq:hamiltonianlaughinlikestate} with the same parameters as in Fig.~\ref{fig:datal240shargedensitywaves}. In the inset, we show the data rescaled in log-log scale. The uncertainties are the standard deviation obtained by computing $A_{\rm CDW}^{(2)}$ with different values of $\Delta L$.}
\label{fig:amplitudechargedensitywavetperp}
\end{figure}

\subsection{Amplitude of the CDW \rev{as a function of $t_\perp$}}
\label{sec:amplitudeofthecdw}
\rev{We now} present our numerical results on the dependence of $A_{\rm CDW}^{(2)}$ on $t_\perp$ (see Sec.~\ref{sec:evaluatingthecdwamplitude}). The data are shown in Fig.~\ref{fig:amplitudechargedensitywavetperp}, using the same parameters as in Fig.~\ref{fig:datal240shargedensitywaves}, and by varying $t_\perp$ over three orders of magnitude, from $t_\perp=10^{-3}$ to $t_\perp=1$. The amplitude of the CDW $A_{\rm CDW}^{(2)}$ is computed \rev{from the spatial pattern of the total density $\langle\sum_y\hat n_{j,y}\rangle$ to which we subtract} the average density, i.e., $A_{\rm CDW}^{(2)}\equiv\max_{j\in I}\langle\sum_y\hat n_{j,y}\rangle-N/L$, where $I=[\Delta L:L-\Delta L]$ is a subregion of the chain to which the boundary sites $\Delta L<L$ are removed, in order to avoid boundary effects. For clarity, the data are reported in log-log scale in the inset. The uncertainties on the data are given by the standard deviation computed by extracting $A_{\rm CDW}^{(2)}$ several times by changing the value of $\Delta L$.

We stress that the bosonization prediction in Eq.~\eqref{cdw2} is valid in the anisotropic limit $t_\perp/t \ll 1$ and in the thermodynamic limit $L\rightarrow\infty$. Differently from what is predicted in Eq.~\eqref{cdw2}, our data for small $t_\perp$ \rev{saturate} to some finite value. This is because, in the limit $t_\perp/t\rightarrow0$, the correlation length diverges, $\xi\rightarrow\infty$. This fact implies that the CDW amplitude is stabilized to its constant bulk value only beyond a number of sites that is sufficiently larger than $\xi$. Explicitly, the condition $L \gg \xi \sim  \left(t_\perp/t \right)^{- \frac{1}{2-X_{{\rm FQH}}}}$, for the smallest values of $t_\perp/t$ that we used and for $X_{{\rm FQH}} \sim 3/2$~\cite{cornfeld2015chiral}, requires $L\gg10^6$. We conclude that, with the limited value of $L=240$ (i.e., of the order of $L=10^2$) that we use in the numerical simulation, we do not have a sufficient range of $L$ in the limit $t_\perp/t\ll1$ to fit the power law of Eq.~\eqref{cdw2}. A much larger value of $L$ would be therefore needed in order to test the scaling as in Eq.~\eqref{cdw2}, but it is unfortunately beyond our numerical possibilities.

\begin{figure}[t]
\centering
\includegraphics[width=8.5cm]{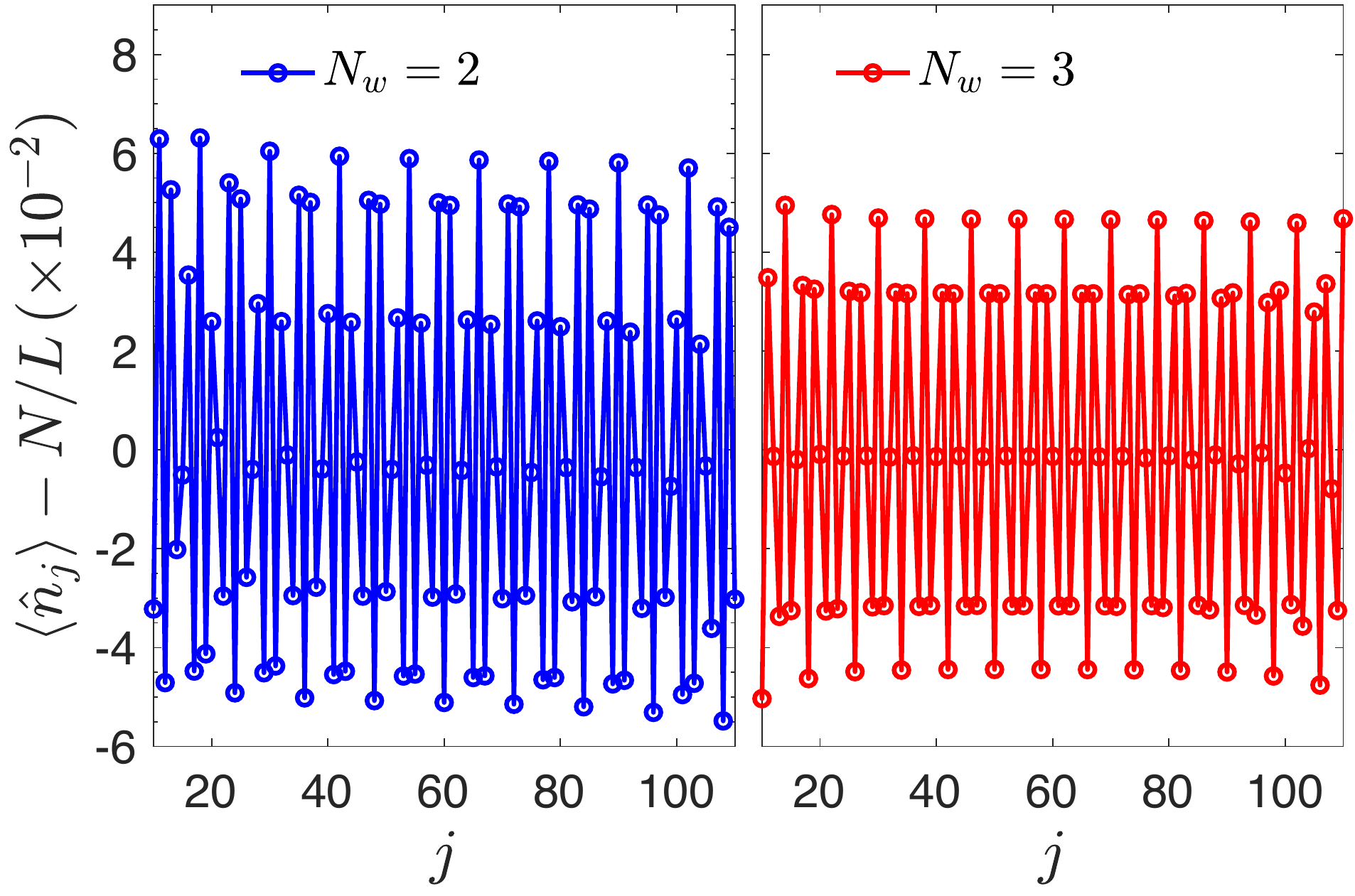}
\caption{\rev{Numerical data for $\langle\hat n_j\rangle-N/L$, where $\hat n_j=\sum_y \hat n_{j,y}$, as a function of $j$, for $j$ around $j=L/2=60$ and not including the chain ends, for $N_w=2$ (left, blue data) and $N_w=3$ (right, red data). The numerical parameters that we use are $t_\perp=10^{-1}\,t$, $V_\perp=30\,t$, $\Phi/\pi\simeq0.832$ and $D_{\rm max}=200$, as in the previous simulations, and $L=120$ in order to reduce the numerical complexity of the problem. We keep $\rho_0$ constant in order to obtain the FQH instability at the same value of $\Phi$ as before, thus we use $N=~50$ ($n=5/12$) for $N_w=2$, and $N=75$ ($n=5/8$) for $N_w=3$. Accordingly, we find in both cases a CDW pattern with spatial period $\lambda=12$ ($N_w=2$) and $\lambda=8$ ($N_w=3$), with decreasing amplitude as a function of $N_w$. We numerically estimate (see also Sec.~\ref{sec:amplitudeofthecdw}) $A_{\rm CDW}^{(2)}\simeq5.9\times10^{-2}$ and $A_{\rm CDW}^{(3)}\simeq4.7\times10^{-2}$.}}
\label{fig:cdwamplitudeasafunctionofnw}
\end{figure}

\subsection{\rev{Amplitude of the CDW for $N_w=2$ and $N_w=3$}}
\label{sec:amplitudecdwnw}
\rev{Before concluding this section, we discuss the dependence of the amplitude of the CDW as the number of wires $N_w$ is increased. The interaction Hamiltonian in Eq.~\eqref{eq:hamiltonianfermionctwolegladder} takes the general form $\hat H_{\rm int}=V_\perp\,\sum_j\sum_{y<y'}\hat n_{j,y}\hat n_{j,y'}$.}

\rev{We show our numerical results in Fig.~\ref{fig:cdwamplitudeasafunctionofnw}. The numerical data of $A_{\rm CDW}^{(N_w)}$ are obtained as discussed in Sec.~\ref{sec:amplitudeofthecdw}. In these simulations, we use $t_\perp=10^{-1}\,t$ and $V_\perp=30\,t$ as before, and keep $\rho_0$ constant in order to have the FQH instability at the same value of $\Phi$ used in the previous sections ($\Phi/\pi\simeq0.832$). The simulations in Fig.~\ref{fig:cdwamplitudeasafunctionofnw} are performed without external chemical potentials, and therefore, because of OBC in the real dimension, the algorithm converges to the CDW with the lowest energy (see also Sec.~\ref{sec:controllingthegroundstatewithexternalchemicalpotentials}).}

\rev{We fix $D_{\rm max}=200$ and, in order to reduce the numerical complexity of the problem, in particular for the simulations with $N_w=3$, for which the hard-core-boson constraint is implemented by using a dimension of the local Hilbert space equal to $8$ on each rung, we keep $L=120$ for both simulations. Accordingly, we use  $N=50$ ($n=5/12$) for $N_w=2$, and $N=75$ ($n=5/8$) for $N_w=3$. As expected from Eq.~\eqref{eq:CDWbasic}, in both cases, we observe a CDW pattern with spatial period $\lambda\propto~1/(N_w\rho_0)$. In particular, on the lattice, we numerically find $\lambda=12$ (for $N_w=2$) and $\lambda=8$ (for $N_w=3$) sufficiently far away from the chain ends.}

\rev{From our numerical result, we observe that $A_{\rm CDW}^{(2)}>~ A_{\rm CDW}^{(3)}$ (in particular, we numerically estimate $ A_{\rm CDW}^{(2)}\simeq5.9\times10^{-2}$ and $A_{\rm CDW}^{(3)}\simeq4.7\times10^{-2}$), which is compatible with Eq.~\eqref{cdwnw}. In order to further corroborate this result, a deeper numerical analysis of the scaling of the CDW amplitude with $N_w$ is needed. In addition to the exponential increase of the local Hilbert space ($2^{N_w}$ on each rung), this may require also the increase of $D_{\rm max}$ in order to ensure the correct convergence of the algorithm. This is for the moment beyond our numerical possibilities, and we leave this task for future work.}

\section{Non-Abelian zero modes in 1D}
\label{se:anyons1D}
\rev{In the previous sections, we analytically and numerically discussed in detail the emergence of a CDW in thin FQH cylinders as a function of system parameters, such as the width $N_w$ and the inter-wire hopping $t_\perp$. Our analytical analysis building on the wire construction approach allowed us to connect the phase of the charge density wave, a notably local order parameter, with the eigenvalue of non-local Wilson loop operators, signifying non-local topological degeneracy. In the thick cylinder limit, the amplitude of the local CDW decays exponentially with $N_w$. Thus, we have explicitly described the crossover between the topological and non-topological regimes of a FQH state with finite dimensions.}
 
\rev{In this section, we extend the discussion to more general geometry, specifically to higher-genus surfaces, on which the FQH state can be embedded. In these general surfaces, additional Wilson loop operators exist and characterize a topological degeneracy in infinite-size limit. However, as any of the dimensions becomes finite and small, based on the previous sections we may deduce a crossover to a local order parameter, where the various quasi-degenerate states can be distinguished by CDWs with different phases. Based on this connection, the goal of this section is to revisit the possibility to realize non-Abelian zero modes in 1D, despite of the apparently forbidding no-go theorems.}

\rev{Specifically, i}n the spirit of Barkeshli \emph{et al.}~\cite{barkeshli2012topological,barkeshli2013twist,barkeshli2014synthetic,barkeshli2015generalized}, we consider extrinsic non-Abelian twist defects, \rev{also known as genons}. We show that these genons in 1D are the pre-topological limit of true anyonic modes occurring in the 2D limit. The splitting of the associated degeneracy can be controlled by the effective width $N_w$ and parametrically by controlling the transverse correlation length~$N^*$.

\subsection{Wilson loops on higher-genus surfaces}
\label{se:simulatinglatticedefects}
\begin{figure}[t]
	\centering	
	\includegraphics[width=8cm]{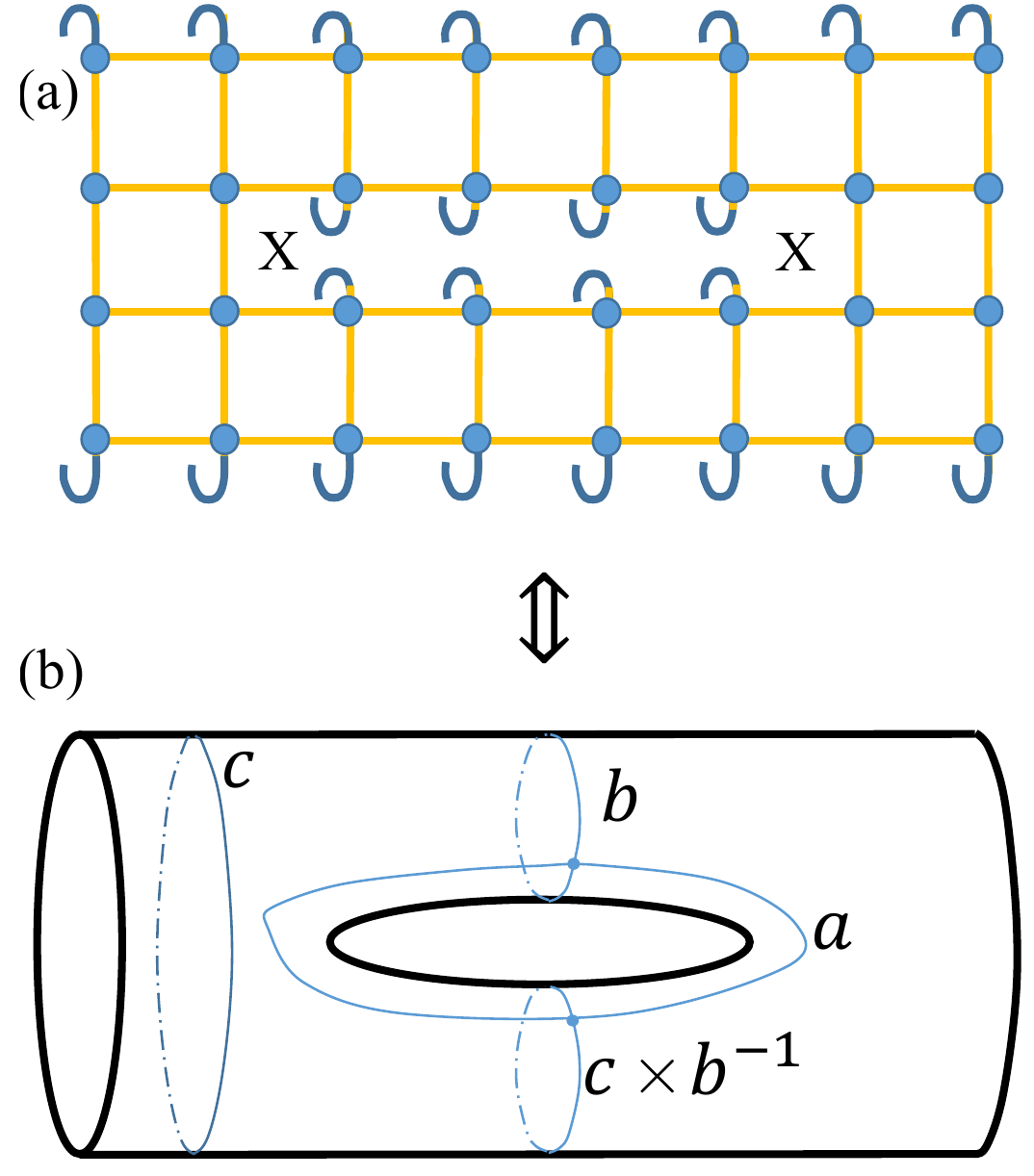}
	\caption{Lattice defects X creating high genus surfaces behave as $\mathbb{Z}_{2q}$ parafermions~\cite{barkeshli2013twist}. The splitting of their associated ground states is exponentially small in lengths of loops $b$ or $c \times b^{-1}$.}
	\label{fg:genus}
\end{figure}

Let us imagine creating lattice defects and test how the pre-topological FQH
state responds. One of the simplest examples of a genon-like topological defect
is illustrated in Fig.~\ref{fg:genus}, panel \textbf{(a)}, where the central
region of a four-leg ladder is transformed into a pair of two-leg ladders.

In the continuum limit shown in Fig.~\ref{fg:genus}, panel \textbf{(b)}, this
would be equivalent to increasing the genus of the manifold by creating an
extra handle. In addition to the loop $c$ winding around the cylinder, we now
have a loop $a$ winding around the new hole, and a loop $b$ circulating around
one of the smaller cylinders forming the handle. Loops $a$ and $b$ intersect at
one point. Consequently, an additional $2q$-fold degeneracy is associated
with this handle, as can be formally seen by constructing Wilson loops $W(a)$
and $W(b)$ and showing that they satisfy the magnetic algebra $W(a) W(b) = W(b)
W(a) \,e^{i\,2 \pi /(2q)}$. However, this degeneracy is not exact in a finite
system.

As explained in the end of Sec.~\ref{se:1D2D}, any small impurity coupling to
the local density will immediately split the degeneracy of the cylinder, adding
a Wilson-loop term $\hat H = -A_c W(c)+{\rm H.c.}$ to the ground state
Hamiltonian. The amplitude of this term is exponentially small in the length of
this loop, $N_w$. Similarly, for any finite 2D manifold such as
Fig.~\ref{fg:genus}, panel \textbf{(b)}, the Hamiltonian acting
within the ground-state subspace is \be \label{eq:Wilsonloop} \hat H_{\rm GS} =
- \sum_{\mathcal{C} } A_\mathcal{C} W(\mathcal{C})+{\rm H.c.} \,\,, \ee where
$\mathcal{C}$  runs over all non-contractible loops.  This Hamiltonian leads to
splitting of the degeneracy by an amount proportional to $A_\mathcal{C}$. For a
rectangular loop of dimensions $L_\mathcal{C} \times N_{\mathcal{C}}$, where
$L_\mathcal{C}$ is a distance along the wires, and $N_\mathcal{C}$ is a
distance perpendicular to the wires, \rev{the amplitude of a Wilson loop in Eq.~\eqref{eq:Wilsonloop} in terms of its length is}
\begin{equation}
A_\mathcal{C} \sim e^{-N_{\mathcal{C}}/N^*} e^{-L_\mathcal{C}/\xi} \,\, .
\label{eq:wildonloopamplitude}
\end{equation}

\rev{Thus, while in the previous sections we have obtained explicitly the Wilson loop operator in Eq.~\eqref{eq:kappa} for the CDW along an infinite cylinder, in this section, we conjecture that any Wilson loop in a general geometry, such as the one in Fig.~\ref{fg:genus}, represents a phase of a CDW in the limit where the length of the loop is small. For example, the eigenvalues of Wilson loops $b$ and $c \times b^{-1}$ represent CDW phases along the individual top and bottom cylinders, respectively. The Wilson loop $a$, in the limit where the hole in Fig.~\ref{fg:genus} is small, represents another CDW pattern. Crucially, the non-commutativity of Wilson loop operators, implies that one can not measure simultaneously these CDWs.}

\subsection{Non-Abelian modes bound to lattice defects in thin cylinders}

\rev{While the entire discussion can be made in terms of the Wilson loop operators, which play a central role in this paper, Wilson loop operators, specifically $W(a)$ and $W(b)$} are formally related to parafermionic
genons~\cite{barkeshli2013twist}. While Wilson loop operators are gauge
invariant, one can construct non-gauge-invariant operators with support near
the point defects [$X$ in Fig.~\ref{fg:genus}, panel \textbf{(a)}] which are
parafermionic operators with quantum dimension
$\sqrt{2q}$~\cite{barkeshli2013twist}. Denoting these parafermionic operators
by $\chi_j$, $j=1,2$, one can \rev{symbolically} write  $W(a) = \chi_1^\dagger \chi_2$.

\rev{We now wish to use our results, specifically Eq.~\eqref{eq:wildonloopamplitude}, to show that the degeneracy associated with the hole in Fig.~\ref{fg:genus} can not be exact, precisely because the system is 1D, but it can be made exponentially exact. With the above formal connection to zero-modes, this will make our point that, despite of the no-go theorems, non-Abelian zero modes with exponential protection can be \textit{de facto} realized in 1D.}

Imagine taking the length of such a quasi-1D system with a hole to infinity,
$L_a \to \infty$. In this case the amplitude $A_a$ of the $a$-loop is
vanishingly small. One may naively deduce that  the two parafermions are
spatially separated and hence topologically protected. However this is not
true, since the system is 1D. Indeed, we have a small Wilson loop $b$, which
does not commute with $W(a)$, whose amplitude in the Hamiltonian is proportional to $e^{-(N_w/2)/N^*}$ and thus it is only suppressed by the \emph{width} of the system. Thus it
will generically appear in the Hamiltonian and split the degeneracy. The
analysis of Sec.~\ref{se:1D2D} shows that the associated states correspond to a
CDW order appearing on the small cylinders forming the handle. On the other
hand, upon increasing $N_w$, but still keeping it finite, one can readily reach
the regime whereby the CDW order is effectively no longer detectable, and hence
the parafermionic zero modes become \textit{de facto} topological.

Envisioning quantum information applications, one could potentially control
$N^*$, which depends on system parameters such as $t_\perp$, thus driving the
system across the topological-nontopological crossover. Quantum information can
then be read in the latter regime whereas it can be stored and manipulated in
the former. We discuss in Appendix~\ref{appendix:parafermions} possible
manipulations with multiple holes.

\rev{The discussion in this section was limited to general arguments, which allowed us to draw generic conclusions. On the other hand, a detailed analytical as well as numerical analysis would be essential to predict specific protocols for manipulations of these genons. A number of important points have remained unexplored, such as quantum superpositions of non-commuting CDWs (eigenstates of non-commuting Wilson loops). We leave this formidable task for future study.}

\section{Conclusions}
\label{sec:conclusions} In this work, we discussed the crossover between the
1D Laughlin-like state and the 2D Laughlin state on a
torus, focusing on flux-ladder setups. This dimensional crossover has been
analyzed by means of a wire construction, specifically, by considering a flux
ladder with $N_w$ wires subjected to an effective gauge field. In the thin
torus limit of $N_w=2$, the bosonic Laughlin-like state at filling factor
$\nu=1/(2q)$ displays $2q$ degenerate ground states that can be locally
distinguished by the local particle density, and are given by CDWs whose
spatial period is related to the total particle density in the system. Using
bosonization arguments, we demonstrated that the amplitude of such CDWs is
exponentially suppressed as the number of legs increases, i.e., when
approaching the 2D topological Laughlin state.

We analyzed in detail the thin torus limit of the bosonic $\nu=1/2$
Laughlin-like state in the two-leg flux ladder. Starting from the chiral
Laughlin-like state studied in previous work, the thin torus geometry was
achieved by including an additional inter-leg hopping with an additional gauge
flux that depends on the particle density, such that a full gap in the
low-energy spectrum of the Laughlin-like state is induced. By means of numerical simulations based on MPS,
we have been able to simulate this thin torus limit of the $\nu=1/2$
Laughlin-like state. By locally controlling the CDW pattern in different
subregions of the ladder, we simulated domain walls between the two (quasi-)degenerate ground states, which allowed us to measure the fractional
charge excitations with charge $|\nu|=1/2$. Using bosonization arguments to
analize the fate of the CDWs in the two-dimensional $N_w\rightarrow\infty$
limit, we interpreted such fractional charge excitations in the two-leg
flux ladder as precursors of topological fractional excitations in the
bosonic $\nu=1/2$ Laughlin state. \rev{We also compared the numerical results of the CDW amplitude for $N_w=2$ and $N_w=3$, and we indeed observed that the CDW amplitude decreases with $N_w$}. Finally, we discussed the possibility of
hosting unprotected non-Abelian zero modes in ladder setups. Such modes are
pre-topological analogues of topologically protected genons, i.e., non-Abelian
twist defects in a bilayer Laughlin state in~2D.

The advantage of focusing on flux ladders stems from the fact that these
systems are of direct relevance and at the nowadays reach in ultra-cold atom
experiments, either employing real or synthetic dimensions. Focusing on the
latter case, the longitudinal direction of the ladder is generated by
counter-propagating lasers that create an optical lattice, which controls the
longitudinal hopping parameter $t$, in which atoms are loaded, and the
transverse (synthetic) dimension is generated by exploiting some internal
atomic quantum numbers. Additional clock or Raman beams are used to induce
transitions between such internal states, where $t_\perp$ and $\Phi$ are
controlled by the intensity of the additional beam and its angle of incidence
relative to the longitudinal direction of the ladder, respectively. Our two-leg
ladder setups employ an additional gauge-flux term to close the Laughlin-like
state on a thin torus that can be realized by a secondary Raman beam with a
different angle of incidence, which depends on the particle density.

Quantum
gas microscopes can provide a single-site high-resolution of the particle
density that, on the one hand, allows to measure the particle density with high
accuracy, and then determine the angle of incidence of the secondary Raman
beams, and, on the other hand, allows to visualize the CDW pattern along the
ladder (see Ref.~\cite{strinati2017laughlin} and references therein), \rev{and therefore measure the fractional excitations}. \rev{Moreover},
one can envision proper engineering of optical superlattices in order to
generate the local chemical potentials that we used, and then control the CDW
pattern along certain portions of the ladder. \rev{Finally, synthetic dimension in cold-atom setups offers a flexible platform to realize ladder configurations with non-trivial topology~\cite{Boada_2015,PhysRevLett.121.150403}, which is obtained by properly engineering the connectivity between different synthetic states. This allows us to reasonably envision the experimental realization of the topological defects discussed in this work.}

In order to further establish this 1D-to-2D crossover, a deeper numerical
analysis extended to the case of many coupled wires, also reproducing the
presence of lattice defects, and/or to very large systems is a highly desirable
goal. This subject is left for future studies.

\begin{acknowledgements}
We thank \rev{Michele Burrello}, Emanuele~G. Dalla~Torre, Leonardo~Mazza, Guido~Pagano, Efrat~Shimshoni, and Lior~Silberman for fruitful discussions. We are grateful to Richard~Berkovits and Davide~Rossini for support. E.~S. and K.~S. were supported by the US-Israel Binational Science Foundation (Grant No.~2016255). M.~C.~S. acknowledges support from the Israel Science Foundations, Grants No.~231/14 and No.~1452/14. S.~S. acknowledges support from~NSERC.
\end{acknowledgements}

\appendix

\section{Evaluating the correlation function in Eq.~(\ref{Eq:ACDW})}
\label{appendix:B}
In this appendix, we report the evaluation of the correlation function in Eq.~\eqref{Eq:ACDW}.

\subsection{$N_w$-leg ladder}
The strongly fluctuating fields $\{\tilde{\theta}_{y\pm\frac{1}{2}}\}$  yield a $(N_w+1)$-point function that decays exponentially at long distance with a typical correlation length $\xi \sim v/\Delta_{\rm gap}$ determined by the inverse gap $\Delta_{\rm gap}$ opened by the relevant FQH Hamiltonian $\hat H_{\rm FQH}$~\cite{cornfeld2015chiral}. To evaluate it we use a simple massive approximation for correlation functions~\cite{giamarchi2003quantum}:
\begin{equation}
\label{eq:expoint}
\prod_{y=1}^{N_w} \left\langle e^{i \sum_j B_{j}^{(y)} \tilde{\theta}_{j, y+\frac{1}{2}} } \right\rangle \! \simeq \! \prod_{y=1}^{N_w} e^{\frac{1}{2} \sum_{i<j}B^{(y)}_i B^{(y)}_j \frac{\sqrt{x_{ij}^2+v^2 \tau_{ij}^2}}{\xi}} \,\, ,
\end{equation}
where we use the notation $\tilde{\theta}_{j, y+\frac{1}{2}} \equiv \tilde{\theta}_{ y+\frac{1}{2}}(x_j, \tau_j)$, $x_{ij}=|x_i-x_j|$ and $\tau_{ij}=|\tau_i-\tau_j|$. Here, $j=0,1,\dots, N_w$ labels the $N_w+1$ space-time points, where $j=0$ corresponds to the bare density operator ($x=0$ and $\tau=0$), and $1 \le j \le N_w$ correspond to the other $N_w$ fields at space-time points $(x_j,\tau_j)$ arising in the perturbative calculation. %In computing Eq.~\eqref{eq:expoint}, we used a long distance approximation for correlation functions, which can be derived by replacing the cosine by a mass term inversely proportional to $\xi$.

To find the $B$'s coefficients in Eq.~\eqref{eq:expoint}, we observe that for $j=0$ we have the operator $e^{-2\,i\,\phi_{y_0}}$, so using Eq.~(\ref{transfREV}) we have
\begin{equation}
B_0^{(y)} = -\delta_{y,y_0-1}\,\frac{1}{2q}+\delta_{y,y_0}\,\frac{1}{2q} \,\, .
\end{equation}
For $1 \le j \le N_w$ we consider the operator $\mathcal{O}^{j \to j+1}_{p_j \to p_j'} (x_j , \tau_j)$. Using Eq.~(\ref{transfREV}), we have
\begin{eqnarray}
\label{Eq:B}
&&B_j^{(y)} = \delta_{y,j}\,\frac{p_j+p_j '}{2q}\nonumber\\
&&\hspace{0.4cm}+\delta_{y,j-1}\left(\frac{1}{2} - \frac{p_j}{2q} \right)+\delta_{y,j+1}\left(-\frac{1}{2} - \frac{p_j '}{2q} \right) \,\, .
\end{eqnarray}
Using Eq.~(\ref{eq:pppcondition}), we can see that $\sum_j B_j^{(y)}~=~0$ for any $y$.

\subsection{Two-leg ladder}
We now focus on the two-leg ladder, $N_w=2$. We consider filling factor $\nu=1/2$, i.e. $q=1$, described by two pairs of conjugate fields:
\begin{equation}
\label{eq:phitheta2leg}
\begin{array}{l}
2\,\tilde{\phi}_{\frac{1}{2}}(x) =\theta_{1} - \theta_{2}+ 2q\,(\phi_{1} + \phi_{2})  \\\\
2\,\tilde{\theta}_{\frac{1}{2}}(x) =\theta_{1} + \theta_{2}+ 2q\,(\phi_{1} - \phi_{2})  \\\\
2\,\tilde{\phi}_{-\frac{1}{2}}(x) =\theta_{2} - \theta_{1}+ 2q\,(\phi_{1} + \phi_{2})  \\\\
2\,\tilde{\theta}_{-\frac{1}{2}}(x) =\theta_{2} + \theta_{1}+ 2q\,(\phi_{2} - \phi_{1}) \,\, ,
\end{array}
\end{equation}
where we denoted by $1/2$ the link between 1 and 2, and by $-1/2$ the other link, and the inverse of this transformation is
\begin{equation}
\label{basistransformation}
\begin{array}{l}
4q\,\phi_{1} = \tilde{\phi}_{-\frac{1}{2}}- \tilde{\theta}_{-\frac{1}{2}} +\tilde{\phi}_{\frac{1}{2}} + \tilde{\theta}_{\frac{1}{2}}  \\\\
2\,\theta_{1} = -\tilde{\phi}_{-\frac{1}{2}}+ \tilde{\theta}_{-\frac{1}{2}} +\tilde{\phi}_{\frac{1}{2}} + \tilde{\theta}_{\frac{1}{2}} \\\\
4q\,\phi_{2}  = \tilde{\phi}_{\frac{1}{2}}- \tilde{\theta}_{\frac{1}{2}} +\tilde{\phi}_{-\frac{1}{2}}  + \tilde{\theta}_{-\frac{1}{2}}\\\\
2\,\theta_{2} = -\tilde{\phi}_{\frac{1}{2}}+ \tilde{\theta}_{\frac{1}{2}} +\tilde{\phi}_{-\frac{1}{2}}  + \tilde{\theta}_{-\frac{1}{2}}  \,\, .
\end{array}
\end{equation}
A special feature of the $N_w=2$ case is that for $p=p'=~0$ the link operators $\mathcal{O}_{p p'}^{y \to y+1}$ involve gapped fields only. Indeed, these link operators contain $\theta_1-\theta_2 = \tilde{\phi}_{\frac{1}{2}} - \tilde{\phi}_{ - \frac{1}{2}}$. Using Eq.~\eqref{eq:pppcondition}, and by taking $y_0=1$ without loss of generality, one has $p_2=p'_1-1$ and $p_2'=p_1$, and then the CDW amplitude Eq.~(\ref{Eq:ACDW}) for the two-leg ladder is
\begin{eqnarray}
&&A_{\rm CDW}^{(2)}\!=\!2\beta_{1,y_0} t_\perp^2 \int dx_1 \,d\tau_1 \,dx_2 \,d\tau_2 \sum'_{p_1,p'_1 }\,C_{p_1,p_1'}^{1,2}\,C_{p_1'-1,p_1}^{2,1} \nonumber\\
&&\!\times\,e^{i [\Phi - i (p_1-p_1')2 \pi \rho_0 ]x_1} e^{i [\Phi - i (p'_1-1 - p_1 )2 \pi \rho_0 ]x_2}\nonumber\\
&&\!\times\,\left\langle e^{-2 i \phi_1(0,0)} e^{i(2p_1 \phi_1-2 p_1' \phi_2)_{x_1,\tau_1}} e^{i[2(p_1'-1) \phi_2-2 p_1  \phi_1]_{x_2,\tau_2}} \right\rangle \,\, . \nonumber\\
\label{AACCDDWW}
\end{eqnarray}
The operator $\mathcal{O}_{0 \to 0}$ is a constant for the two-leg ladder, and should not be included, hence $(p_1,p_1 ') \ne (0,0)$ and $(p_1'-1, p_1) \ne (0,0)$. Similarly $(p_1,p_1') \ne (1,-1)$ or $(-1,1)$ which are the two FQH operators.
%In the correlator we omitted $\theta_1 - \theta_2$ that is a constant.
Using Eqs.~\eqref{eq:expoint} and~\eqref{basistransformation}, one has
\begin{equation}
\begin{array}{l}
B_0^{\frac{1}{2}} = \cfrac{1}{2q} \qquad B_0^{-\frac{1}{2}} =-\cfrac{1}{2q}\\
B_1^{\frac{1}{2}} =- B_1^{-\frac{1}{2}} =   \cfrac{p_1+p_1 '}{2q} \\
B_2^{\frac{1}{2}} =- B_2^{-\frac{1}{2}} =   -\cfrac{p_1+p_1 '-1}{2q}
\end{array} \,\, ,
\end{equation}
satisfying $\sum_j B_j^{(y)}=0$ for any $y$. Thus, at filling factor $\nu=1/2$, one has $\Phi=4\pi\rho_0$, and then using Eq.~(\ref{eq:expoint}), we obtain the integral given in Eq.~(\ref{xi4}) in the main text.

\section{Evaluation of the strongly oscillating integral $I(\kappa)$ in Eq.~\eqref{IPP}}
\label{appendix:evaluationoftheintegral}
In this appendix, we report the explicit calculation of the integral determining the amplitude of the CDW in Eq.~\eqref{IPP}. Consider the integral in Eq.~(\ref{IPP}) in the limit of large $\kappa$. The goal of this appendix is to show that it decays as $1/\kappa^5$. Thus, $A_{\rm CDW}^{(2)} \sim (t_\perp/t)^2(\Delta_{\rm gap}/t) $. Also, for the $N_w>2 $ generalization of this integral (with a prefactor $\xi^{2 N_w}$ pulled out), we will obtain a $1/\kappa^{2 N_w+1}$ decay so that  $A_{\rm CDW}^{(N_w)}\sim (t_\perp/t)^{N_w} (\Delta_{\rm gap}/t)$.

The simplest way to evaluate strongly oscillatory integrals is integration by parts. To illustrate this, consider the integral
\be
\label{intbyparts}
\mathcal{I}[f(x),k]=\!\int_0^\infty e^{i k x} f(x)\,dx=\!\sum_{m=0}^\infty \left(\frac{1}{i k} \right)^{m+1} \!\!f^{(m)}(0) \,\, .
\ee
Here, $e^{i k x}$ is the strongly oscillating function in the limit of large $k$, and $f(x)$ is some smooth function. The expansion involves the value of $f^{(0)}(x)=f(x)$ and its derivatives $f^{(m)}(x)=\partial_x^m f(x)$ at $x=0$. To derive this \rev{expansion}, one repeatedly writes the strongly oscillating function as $e^{ik x} = {(ik)}^{-1}(d\,e^{i k x}/dx)$ and integrates by parts. This can be readily checked for simple functions such as $f(x) = x^p e^{-x}$. Note that if $f(x)$ is continuous and finite for $x \in (-\infty  , \infty)$ then the expansion of the integral as $k \to 0$ can be non-analytical; for example for $f(x) = 1/(1+x^2)$, we have $\mathcal{I}[f(x),k] \propto e^{-k}$ which is not analytic at $k \to \infty$.
%%%%%
\begin{figure} [t]
	\centering	
	\includegraphics[width=5cm]{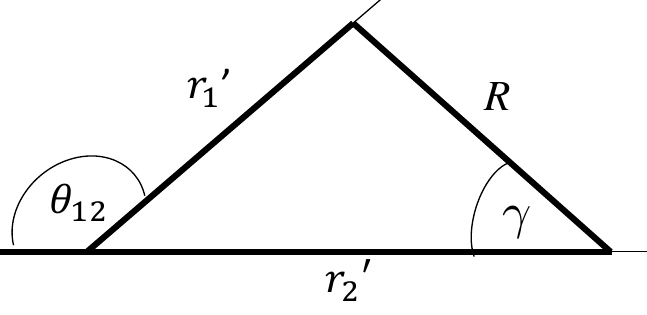}
	\caption{Illustration of change of variables in Eq.~(\ref{changevari}).}
	\label{fgtriangle}
\end{figure}
%%%%%
We now bring our integral $I_{p_1, p_1 '}(\rho_0 \xi)$ to a form where we can use the integration by parts method with respect to a single semi-infinite variable.

\subsection{Performing analytically one integral}
Going to polar coordinates $x_1 = r_1 \cos(\theta_1)$, $t_1 = r_1 \sin(\theta_1)$ and similarly for $x_2$ and $t_2$, and using $r_{12}=\sqrt{[r_1 \sin (\theta_{12})]^2+[r_2-r_1 \cos (\theta_{12})]^2}$ with $\theta_{12} = \theta_1 - \theta_2$, we have
\bea
&&I_{p_1, p_1 '}(\kappa)=\int_0^\infty dr_1\,r_1 \int_0^\infty dr_2\,r_2 \int_0^{2\pi} d\theta_1 \int_0^{2\pi} d\theta_2 \nonumber \\
&&\hspace{0.4cm}\times\,e^{2 \pi i \kappa (r_1 \cos(\theta_1) [2 - (p_1 - p_1 ')]+ r_2 \cos(\theta_2) [2 - (p_1 ' -1 - p_1)])} \nonumber \\
&&\hspace{0.4cm}\times\,e^{-r_{1}\frac{p_1 + p_1 '}{(2q)^2}} e^{r_{2} \frac{p_1 + p_1 ' -1}{(2q)^2}} e^{-r_{12} \frac{(p_1 + p_1 ')(p_1 + p_1 '-1)}{(2q)^2}} \,\, .
\eea
Now we change the angular variables to $\alpha = (\theta_1+\theta_2)/2$ and $\theta_{12} = \theta_1 - \theta_2$. One has for the angular part
\be
\int_0^{2\pi} d\theta_1 \int_0^{2\pi} d\theta_2=\int_0^{2\pi} d \alpha \int_0^{2\pi} d\theta_{12} \,\, .
\ee
We will first perform analytically the $\alpha$ integral. Only the oscillating factor depends on $\alpha$. Using trigonometric identities and $\int_0^{2 \pi} d \alpha\,e^{i a \cos \alpha} = 2 \pi J_0 (a)$, where $J_0(\cdot)$ is the Bessel function of the first kind~\cite{abramowitz1965handbook}, we have
\bea
\int_0^{2 \pi} && d \alpha\,e^{2 \pi i \kappa (r_1 \cos(\theta_1) [2 - (p_1 - p_1 ')]+ r_2 \cos(\theta_2) [2 - (p_1 ' -1 - p_1)])}  \nonumber \\
&&=2 \pi  J_0 \left(\kappa \sqrt{{r_1'}^2+{r_2'}^2+2 {r_1'} r_2' \cos(\theta_{12})} \right) \,\, ,
\eea
where $r_1'=2 \pi  |2-(p_1-p_1')| r_1 $ and $r_2'=2 \pi   |2-(p_1'-1-p_1)| r_2$.
Thus we are left with a three-dimensional integral
\bea
I_{p_1, p_1 '}(\kappa)=\int_0^{2\pi} d\theta_{12}  \int_0^\infty dr_1\,r_1 \int_0^\infty dr_2\,r_2  \nonumber \\  2 \pi J_0 \left(\kappa \sqrt{{r_1'}^2+{r_2'}^2+ 2{r_1'} r_2' \cos(\theta_{12})}\right) \nonumber \\
	e^{-r_{1}\frac{p_1 + p_1 '}{(2q)^2}} e^{r_{2} \frac{p_1 + p_1 ' -1}{(2q)^2}} e^{-r_{12} \frac{(p_1 + p_1 ')(p_1 + p_1 '-1)}{(2q)^2}} \,\, ,
\eea
whose evaluation is discussed in the next section.

\subsection{Change of variables}
For fixed $\theta_{12}$, we can think of $r_1'$ and $r_2'$ as the lengths of two vectors $\vec{r}_1, \vec{r}_2$ emanating from the origin along two rays with angle $\theta_{12}$. Then, the argument of the Bessel function is $\kappa R$, where $R=\sqrt{{r_1'}^2+{r_2'}^2+ 2{r_1'} r_2' \cos(\theta_{12})}$ is the distance between the heads of these two vectors, i.e. the length of $\vec{r}_1 - \vec{r}_2$. It is more convenient to change variables of integration $r_1',r_2'$ into $R$ and $\gamma$, where $\gamma$ is the angle between the sides of lengths $r_1'$ and $R$ on this triangle, see Fig.~\ref{fgtriangle}. Then
\bea
\label{changevari}
r_1'= \frac{R \sin(\gamma)}{\sin(\theta_{12})} \qquad r_2'=\frac{R \sin(\gamma)}{\tan(\theta_{12})}+R \cos(\gamma) \,\, .
\eea
Including the Jacobian of this transformation
\begin{equation}
\frac{dr_1'}{d\gamma}\frac{dr_2'}{d R} - \frac{dr_2'}{d\gamma}\frac{dr_1'}{d R}  = \frac{R}{\sin(\theta_{12})} \,\, ,
\end{equation}
we have
\bea
\label{eq:33}
\int_0^\infty d r_1 ' \,r_1' \int_0^\infty dr_2' \,r_2 ' = \int_0^\infty d R \,R^3 \int_0^{\pi - \theta_{12}} d \gamma  \nonumber \\
 \frac{1}{\sin(\theta_{12})} \frac{ \sin(\gamma)}{\sin(\theta_{12})} \left[ \frac{ \sin(\gamma)}{\tan(\theta_{12})}+ \cos(\gamma) \right] \,\, .
\eea

\subsection{Expanding the strongly oscillating Bessel function}
Similar to Eq.~(\ref{intbyparts}), we can consider
\bea
\label{intbypartsBESSEL}
&&\mathcal{J}[f(x),k] = \int_0^\infty J_0(k x) f(x) \,dx \nonumber \\
&&= \frac{1}{k} f(0)- \frac{1}{2}\frac{1}{k^3} f''(0)+\frac{3}{8} \frac{1}{k^5} f''''(0)+\ldots \,\, ,
\eea
which is derived in the same way. To obtain the first term in this expansion one replaces $J_0(x) = \partial_x [\int^x dx' J_0(x')~+~c_1]$, and chooses the constant $c_1$ such that the resulting function decays at infinity. This procedure is repeated to all orders, obtaining a different $c_m$ at the $m$-th order, and the expansion coefficients are the resulting values of the $\{c_m\}$. One can check this expansion for analytically solvable integrals e.g. for $f(x) = e^{-x}$. Using this expansion, together with the form Eq.~(\ref{eq:33}) we can immediately determine the leading power law decay of our integral with $\kappa$. The $R^3$ factor implies that the leading order contribution is the $\kappa^{-5}$ term in Eq.~(\ref{intbypartsBESSEL}).

\subsection{Generalization to $N_w$ wires}
Consider the general expression Eq.~(\ref{Eq:ACDW}) for $A_{\rm CDW}^{(N_w)}$. Using the exponentially decaying approximation for the correlation functions, Eq.~(\ref{eq:expoint}), we may repeat the procedure leading to Eq.~(\ref{xi4}), $A_{\rm CDW}^{(N_w)} = \rho_0 (t_\perp/t)^{N_w} \xi^{2 N_w} I^{(N_w)}(\kappa)$.
In polar coordinates the measure of the integral $I^{(N_w)}$, which is the direct $N_w>2$ generalization of Eq.~(\ref{IPP}), is of the form
\be
\label{intmeasure}
\prod_{i=1}^{N_w} \int dx_i \,d\tau_i=  \prod_{i=1}^{N_w} \left[ \int_0^{2\pi} d\theta_i \int_0^\infty dr_i \,r_i\right] \,\, .
\ee
As before we define a global angle $\alpha = (\sum_{i=1}^{N_w} \theta_i)/N_w$, and $N_w - 1$ additional relative angles e.g. $\delta_{i} = \theta_{i+1}-\theta_{i}$, where $i=1,\dots, N_w-1$. We can start by the $\alpha$ integral. As above the only $\alpha$-dependent factor is the oscillating function
\bea
e^{i \kappa \sum_i x_i'}=e^{i \kappa \sum_i r_i' \cos \theta_i},
\eea
with $x_i'=x_i A_i$ with coefficients $A_i$. One can perform the $\alpha$ integral and generate a Bessel function whose coefficient contains $\kappa$. For example for $N_w=3$ one obtains
\bea
\int_0^{2 \pi} d \alpha e^{i \kappa \sum_i r_i' \cos \theta_i} = 2 \pi J_0[\kappa R],
\eea
where $R$ is given by
\begin{widetext}
\begin{equation}
R^2\!=\!{\left[r_1' \cos(2 \delta_1+\delta_2)\!+\!r_2' \cos (2 \delta_1 - \delta_2)\!+\!r_3 ' \cos(2 \delta_1+ 2 \delta_2)\right]}^2\!+\!{\left[-r_1 ' \sin (2 \delta_1 + \delta_2)+r_2 ' \sin(2 \delta_1 - \delta_2 )\!+\!r_3' \sin(2 \delta_2 + 2 \delta_2)\right]}^2 \,\, .
\end{equation}
\end{widetext}
One may provide a cuboid interpretation of this $R$ as a 3D generalization of Fig.~\ref{fgtriangle}.
We may change variables, to include $R$ as the only length, the set of $N_w-1$ variables $\delta_i$ (similar to $\theta_{12}$), and $N_w-1$ additional angular variables (similar to $\alpha$). By dimensional analysis the integral measure in Eq.~(\ref{intmeasure}) depends on $R$ as $\int_0^{\infty} dR\,R^{2 N_w-1} J_0(R \kappa) F(R)$. Here $F(R)$ is the result of doing all the angular variables over the various exponential factors. Generalizing the expansion Eq.~(\ref{intbypartsBESSEL}), we see that $\int_0^\infty dR\,J_0(\kappa R) f(R) = \sum_{m=0}^\infty c_m \,{\kappa}^{-(1+2m)}\,f^{(2m)}$ with coefficients $c_m$ (specifically $c_0=1, c_1=-1/2, c_3=3/8$). From the $R^{2 N_w-1}$ dependence, we see that the leading non-vanishing derivative is the $2N_w-1$ one, hence we get the leading contribution $m=N_w$ from this series, $I^{(N_w)}(\kappa) \propto \kappa^{-(1+2N_w)}$. Thus, the calculation leads to the result $A_{\rm CDW}^{(N_w)} \propto (t_\perp/t)^{N_w+\frac{1}{2-X_{\rm FQH}}}$.

\begin{figure} [t]
	\centering	
	\includegraphics[width=9cm]{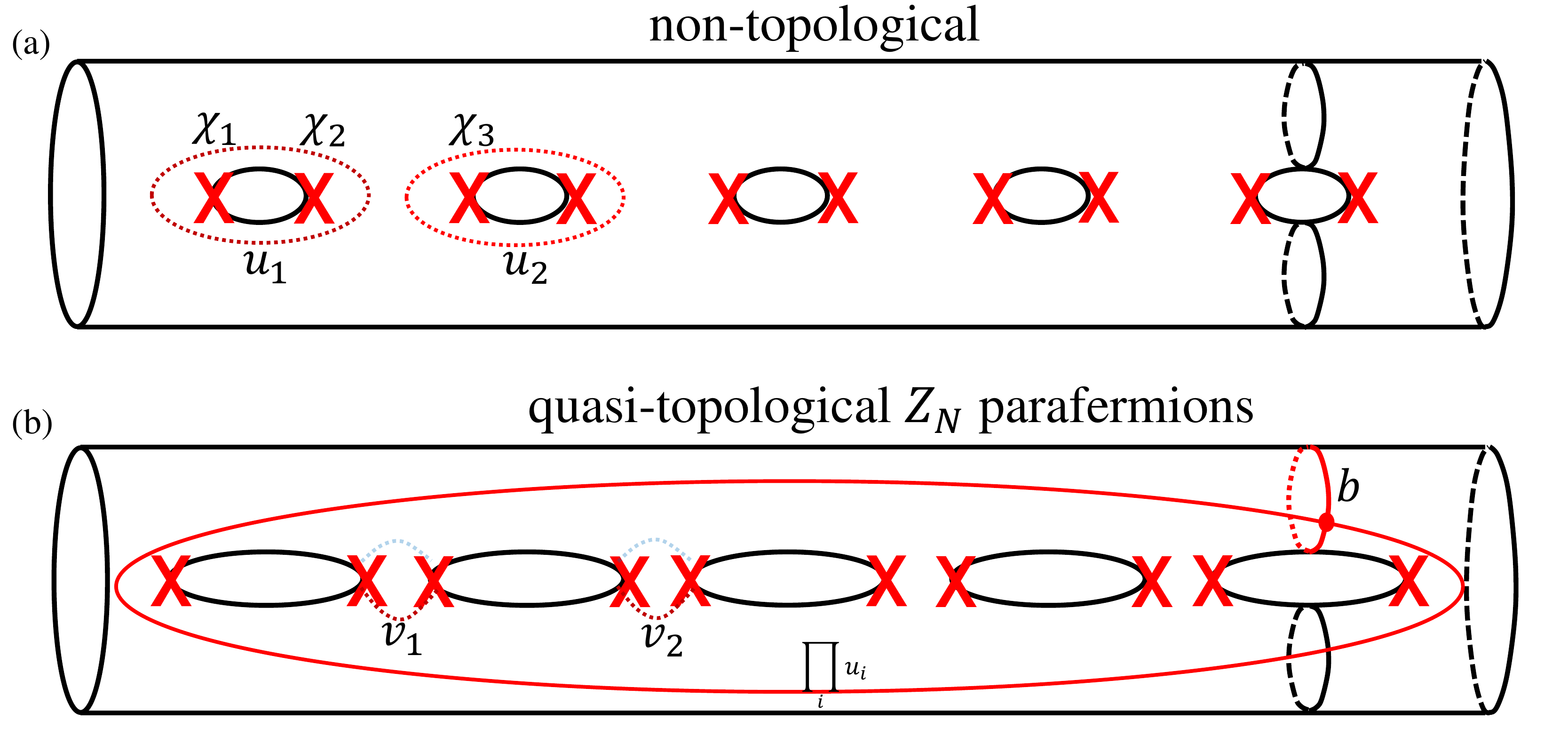}
	\caption{Chain of lattice defects. Each defect carries non-protected $2q-$parafermions. The degeneracy of neighboring \rev{parafermions} is lifted due to Wilson loops generated by local perturbations. The \rev{Hamiltonian} of the two cases is dominated by the shortest loops shown as dashed lines. Both phases are gapped and the extensive ground state degeneracy is removed by the loops. The quantum phase represented at the bottom contains a pre-topological degeneracy associated with the edge parafermions.  The Wilson loops associated with the edge \rev{parafermions} are (i) the loop winding around the entire chain $\prod_i u_i$ which is suppressed as $e^{-L/\xi}$, (ii) but also the loop $b$ controlled by the width $N_w/2$. }
	\label{fg:genus1}
\end{figure}

\section{Smeared density}
\label{app:smeareddensity}
In this appendix, we describe the procedure that we used to compute the fractional charge in Fig.~\ref{fig:domainwallformationscheme2}. Since the charge density field $\phi_y(x)$ does not capture fluctuations on the microscopic scale, e.g., at the level on the single site on the lattice, one can define a smeared density as $n_s(x)=\int dy\,K(x-y)\,n(y)$~\cite{ashcroft2005solid}, where $K(x-y)$ is some kernel normalized such that $\int dx\,K(x-y)=1$, so that $\int dx\,n_s(x)=\int dy\,n(y)=N$. For example, in the continuum case, a Gaussian kernel $K(x)=e^{-x^2/2\sigma^2}/\sqrt{2\pi\sigma^2}$ can work. On a lattice, one has to rewrite the smeared density as
\begin{equation}
n_{s,j}=\sum_hK_{j,h}\,n_h \,\, ,
\label{eq:smeafeddensity2}
\end{equation}
where, in the case of Gaussian kernel, the normalization factor is given in terms of the so-called Jacobian elliptic theta function $\vartheta_3(z,q)$~\cite{abramowitz1965handbook}:
\begin{equation}
K_{j,h}=\frac{e^{-{(j-h)}^2/(2\sigma^2)}}{\vartheta_3(0,e^{-1/2\sigma^2})} \,\, .
\label{eq:kerneldiscretesum}
\end{equation}
In a physical situation, $\sigma$ can be equal to some unit cells. When the
chain has boundaries, i.e., $j,h\in[1:L]$, where $L$ is the chain length, one
has to be careful that the range of $j$ and $h$ has to be extended by some
$2L_0$ auxiliary sites, in order to ensure the correct normalization, i.e.,
$\sum_{j=-L_0}^{L+L_0}n_{s,j}=N$, using the fact that $n_h=0$ for all
$h\notin~[1:L]$, because otherwise the condition $\sum_jK_{j,h}=1$ can not be
fulfilled (the ``violation of the conservation of the number of particles'' on
a chain with sharp boundaries is an artifact of the smearing procedure). This
will of course cause the smeared density in Eq.~\eqref{eq:smeafeddensity2} to
be not a sharp function that goes to zero at the edges, but some nonzero
residual density will be found also for some sites away from the physical edges
of the chain because of the nonlocal nature of the smearing procedure.

\section{1D parafermion chain}
\label{appendix:parafermions}

In this appendix we note that one can generalize the geometry in Fig.~\ref{fg:genus}, panel \textbf{(b)}, to multiple holes, see Fig.~(\ref{fg:genus1}). This realizes an array of parafermions $\chi_i$ where $ \chi_{2i-1}^\dagger \chi_{2i} = W(u_i)$ are the Wilson loops $u_i$ shown in Fig.~\ref{fg:genus1}, panel \textbf{(a)}, and $ \chi_{2i}^\dagger \chi_{2i+1} = W(v_i)$ are Wilson loops $v_i$ shown in Fig.~\ref{fg:genus1}, panel \textbf{(b)}.  One can then choose the dimensions of the holes and the spacing between the holes, such as to control the Hamiltonian Eq.~(\ref{eq:Wilsonloop}) and stabilize various states and phases of parafermions. While in the phase in Fig.~\ref{fg:genus1}, panel \textbf{(a)}, the parafermions are coupled in pairs, due to the dominating $u$-loops, a pair of edge parafermions are left in Fig.~\ref{fg:genus1}, panel \textbf{(b)}, dominated by the $v$-loops. These edge parafermions are non-topological due to the global loop $\prod_i u_i$, which, while being exponentially suppressed with the system size, it does not commute with the small loop $b$ exactly as in Fig.~\ref{fg:genus}, panel \textbf{(b)}. Thus the splitting is actually controlled by $N_w/2$.

\bibliographystyle{apsrev4-1}
%\bibliography{TaoThoulessREFS}


\end{document}